\documentstyle[prl,aps,epsfig,multicol]{revtex}
\newcommand{\beq}{\begin{equation}}
\newcommand{\eeq}{\end{equation}}

\begin{document}

\title{Nanometer scale period sinusoidal atom gratings produced by a Stern-Gerlach
beam splitter}
\author{B. Dubetsky and G. Raithel}

\address{Michigan Center for Theoretical Physics and Physics Department, University\\
of Michigan, Ann Arbor, MI 48109-1120}

\date{\today}
\maketitle

\begin{abstract}
An atom interferometer based on a Stern-Gerlach beam splitter is proposed.
Atom scattering from a combination of magnetic quadrupole and homogeneous
magnetic fields is considered. Using Raman transitions, atoms are coherently
excited into and de-excited from sublevels having nonzero magnetic quantum
numbers. The spatial regions in which the atoms are in such sublevels are
small and have magnetic fields designed to have constant gradients.
Therefore, the atoms experience position-independent accelerations, and the
aberration of the coherently separated and recombined atomic beams remains
small. We find that because of these properties it is possible to envision
an apparatus producing atomic density gratings with nm-scale periods and
large contrasts over $10-100$ $\mu $m. We use a new method of describing the
atomic interaction with a pulsed spatially homogeneous field. In our
detailed analysis, we calculate corrections caused by the non-linear part of
the potential and the finite value of the de-Broglie wave length. The
chromatic aberration and the effects of an angular beam divergence are
analyzed, and optimal conditions for an experimental demonstration of the
technique are identified.
\end{abstract}

\pacs{03.75.-b 03.75.Dg 39.10.+j 32.80.Wr }

\begin{multicols}{2}
\section{Introduction}

An important application of atom interference \cite{c1} is the production of
a periodic spatial profile of the atomic density. When an atomic beam
propagating along the $x$ axis passes through a system of counterpropagating
resonant optical fields having wavelength $\lambda $ or through a
microfabricated structure having period $\lambda /2$ directed along the $z$
(transverse) axis, an initial atomic state having transverse momentum $p$
splits into a series of states having momenta $p+n\hbar k,$ where $n$ are
integers and $k=2\pi /\lambda .$ Interference between these momentum states
results in an atomic density pattern, referred to as a grating, having a
period $\lambda /2.$ A detailed review and bibliography of the different
regimes for producing atomic gratings can be found in our recent article 
\cite{c2}.

For the purpose of this paper, it is important to underline that gratings
with a sinusoidal density profile and nanometer scale period $\lambda
_{g}\ll \lambda ,$ say 
\begin{equation}
\lambda _{g}\sim 10-100\,nm,  \label{i1}
\end{equation}
are of particular interest. One method to achieve this goal is to use a
large angle beam splitter (LABS), which splits the initial atomic state into
two states having momenta $p\pm \Delta p,$ where 
\begin{equation}
\Delta p=\pi \hbar /\lambda _{g}.  \label{i2}
\end{equation}
It was expected that triangular potentials, which one can produce with some
accuracy using a strong standing wave field \cite{c3}, a magneto-optical
scheme \cite{c4}, or bichromatic fields \cite{c5}, can work as a LABS.
However, we recently showed \cite{c2,c21} that, asymptotically, the atom
density profile scattered from this LABS becomes a spatially inhomogeneous
sinusoid with a period of order $\lambda _{g}$ superimposed on sharp density
peaks, separated by $\lambda /2.$ The undesired inhomogeneity results from
the splitting of the atomic state into two {\sl groups} of momentum states
rather than into two well defined momentum states.

In this article we propose a LABS based on the Stern-Gerlach beam splitter 
\cite{c6,c7}. Our main objective is to test the idea of using this LABS to
create sinusoidal atomic gratings with periods much smaller than the optical
wavelength but with coherence lengths much larger than the optical
wavelength.


The article is arranged as follows. In the next Section, we outline the
principles of operation of the proposed devices. In our detailed
calculations, we then consider atomic scattering from a finite thickness
layer of a linear potential (Section III). Corrections for weak and strong
acceleration are evaluated in Sections IV and V. The appendix \ref{atom} is
devoted to calculations of the Raman transitions between Zeeman sublevels.
We summarize our results in Sec.~VI, and discuss their applicability to a
beam of $^{\text{87}}$Rb atoms.

\section{Principle of operation}

\begin{figure}
\begin{minipage}{0.99\linewidth}
\begin{center}
\epsfxsize=.95\linewidth \epsfbox{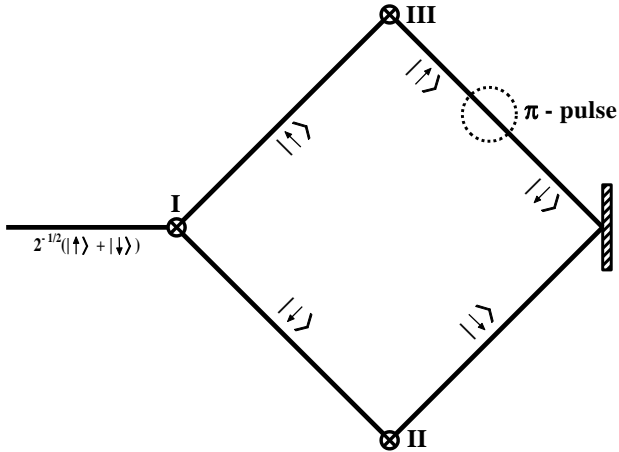}
\end{center}
\end{minipage}
\begin{minipage}{0.99\linewidth} \caption{Principal scheme of the atom interferometer based on the
Stern-Gerlach beam-splitter. 
\label{sg}}
\end{minipage}
\end{figure}

From a naive point of view it seems impossible to use Stern-Gerlach beam
splitters for atom interferometry, since such beam splitters produce
inerferometer arms corresponding to different Zeeman sublevels. These arms
could not interfere. This problem can be avoided as follows (see Fig. \ref
{sg})

In Fig.~\ref{sg} we show the principle of the atom interferometer of
interest. If an atom having angular moment $G=1/2$ is polarized along the $x$
or $y$ axis, then after splitting by the Stern-Gerlach magnet I, the two
arms of the interferometer contain atoms in orthogonal but {\em mutually
coherent}{\bf \ }states having Zeeman quantum numbers $m=\pm 1/2$,
respectively. Magnets II and III reflect the atomic beams in the two
interferometer arms, and recombine them. Before the recombination, the
internal state in one arm of the interferometer is flipped by applying an
RF-induced or optically induced $\pi $-pulse in a spatially homogeneous
magnetic field. All atoms arrive in the recombination region in the same
internal state, and interference occurs.

An ideal sinusoidal grating would arise if an incident plane atomic
wavefunction is split into two coherent components in a spatially
homogeneous magnetic field gradient. Evidently, a realistic Stern-Gerlach
magnet, characterized by a number of higher order spatial derivatives of the
magnetic field, does not satisfy this requirement. To realize a magnetic
field with constant field gradient, we consider the superposition of a
homogeneous bias field ${\bf B}_{s}$ directed along $z$ axis and a
quadrupole magnetic field produced by four 
(anti-)parallel currents propagating in the $\pm y$-directions. This
combination allows us to realize a small volume around the center of the
quadrupole field in which the field gradient is approximately constant, with
no zero of the magnetic field being present. Zeroes of the magnetic field
need to be avoided in order to prevent non-adiabatic spin flips (Majorana
transitions). Since there are no zeroes of $B$, an atom in a magnetic
sublevel $m$ with respect to a quantization axis identical to the $B$-field
direction adiabatically follows direction changes of the $B$-field along the
atom's trajectory, and will always remain in that sublevel. The bias field
also allowed us to generate potential with a dominant linear term and small
nonlinear corrections. The potential for the atom's center-of-mass motion is
then given by $V({\bf r})=g\mu _{B}m\left| {\bf B}({\bf r})\right| $. Atoms
prepared in the $m=0$ sublevel do not interact with the field at all.

Based on the preceding considerations, we can now outline how we realize
Stern-Gerlach beam splitters with small aberration. We assume that the above
described combination of $B$-fields produces a region of almost constant
field gradient centered at the origin. A plane matter wave in state $m=0$
propagates in the $+x$-direction. The wave is not refracted or diffracted by
magnetic-dipole forces while it traverses the fringe fields of the magnets.
External coupling fields are applied in a narrow plane located at $x=-d,$
splitting the wave into a coherent superposition of two waves in different
(relevant) Zeeman sublevels (e. g. $m=\pm 1),$. The spin components of the
wave experience an acceleration $\propto m$, and acquire a differential
transverse momentum change $\Delta p.$ A second set of coupling fields in a
plane at $x=+d$ return the accelerated atoms into the state $m=0$ and
terminate the interaction with the magnetic field. As a result, the system
coherently splits an atomic plane wave entering in a single magnetic
sublevel into two momentum components exiting in the same magnetic sublevel.
aberration effects, i.e. unwanted curvatures in the phase fronts of the
outgoing waves, are minimized by the fact that all magnetic acceleration is
localized to a small region $-d<x<d$, in which the field is not strongly
contaminated with higher-order multipole terms (``fringe fields'').

\begin{figure}
\begin{minipage}{0.99\linewidth}
\begin{center}
\epsfxsize=.95\linewidth \epsfbox{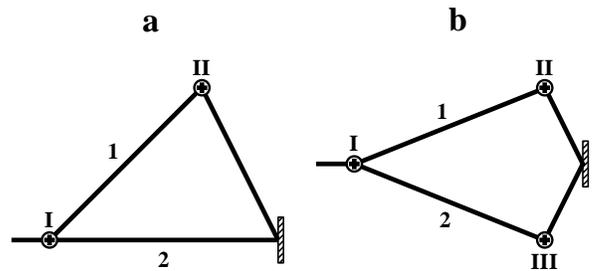}
\end{center}
\end{minipage}
\begin{minipage}{0.99\linewidth} \caption{Schemes to create atomic gratings with (a) asymmetric and (b)
symmetric atom interferometers. I, II, and III are beam-splitters, whose
possible layout is shown in Fig. \ref{f02}. 
\label{f03}}
\end{minipage}
\end{figure}

If the momentum states overlap spatially, an atomic grating will form. The
splitting between Zeeman sublevels caused by the external coupling fields
determines the momentum space distribution and the properties of the atom
grating in the detection plane. The grating phase is determined by the
difference between the atomic wave functions phases acquired along the two
arms of the interferometer. Owing to the phase sensitivity to the atom
velocity and the magnetic field instability, the grating can be washed out,
or one has to require that the atomic beam and field characteristics must be
beyond the current state-of-art. One has to choose a splitting scheme and
interferometer geometry that minimizes this sensitivity. Two types of
interferometers are shown in Fig. \ref{f03}.
 In the triangular
interferometer (Fig. \ref{f03}a) atoms in arms 1 and 2 have different
kinetic energy The phase, associated with this difference, the so called
Talbot phase, leads to periodic oscillations of the density distribution 
\cite{mTlbt}, initially discovered for light \cite{c71} and also observed in
an atom interferometer \cite{oTlbt}. The Talbot phase plays a critical role
in time-domain atom interferometry \cite{c72}, where it has been used for
precise recoil frequency measurements \cite{c73}. For interferometers in the
spatial domain, the Talbot phase degrades the atom grating, and one should
prefer a symmetric interferometer (Fig. \ref{f03}b) that produces no phase
difference.

To produce a symmetric interferometer one first has to split an initial
atomic state $\left| m=0\right\rangle $ symmetrically between, for example,
states $\left| m=\pm 1\right\rangle .$ Co-propagating cross-polarized
optical waves can produce two-quantum transitions between Zeeman sublevels
via excited state manifold to create ground state interferometers. Our
calculations show that a magnetic field in the acceleration zone can lead to
a Zeeman splitting $\omega _{Z}$ larger than the inverse interaction time $%
\tau _{i}$, 
\begin{equation}
\omega _{Z}\tau _{i}\gg 1,  \label{0}
\end{equation}
and effective coupling occurs only if one superposes two waves at
frequencies $\Omega $ and $\Omega +\omega _{Z}$ needed for a two-quantum
resonances. Next, the Zeeman splitting is typically comparable with the
excited state hyperfine splitting $\omega _{hf}$, so that, rigorously
speaking, to obtain two-quantum transition amplitudes one has to know the
excited state manifold structure for a given magnetic field. The situation
simplifies if the detuning $\Delta $ between the waves' frequencies and the
ground-excited state transition frequency is larger than both the Zeeman and
hyperfine splittings, i.e. 
\begin{equation}
\left| \Delta \right| \gg \max \left\{ \omega _{Z},\omega _{hf}\right\} .
\label{1}
\end{equation}

\begin{figure}
\begin{minipage}{0.99\linewidth}
\begin{center}
\epsfxsize=.95\linewidth \epsfbox{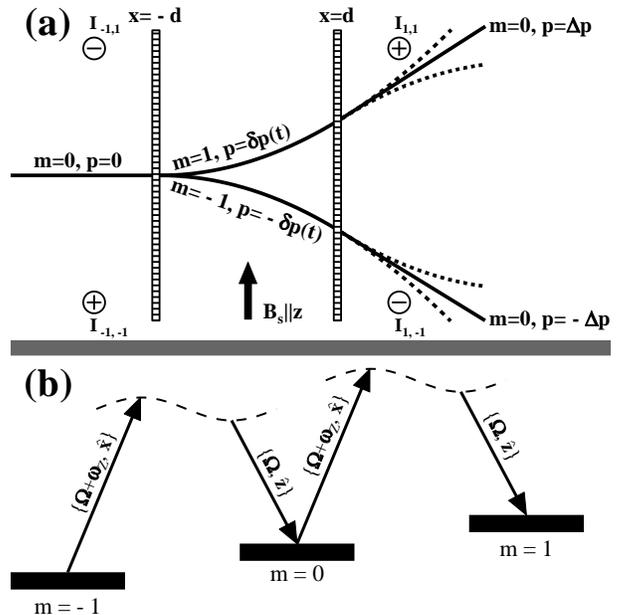}
\end{center}
\end{minipage}
\begin{minipage}{0.99\linewidth} \caption{(a) Scheme of the beam-splitter. The beam splitter involves four
currents $I_{ij}$ comprising a magnetic quadrupole, a homogeneous bias
magnetic field ${\bf B}_{s},$ two thin layers of optical fields located at $%
x=\pm d.$ Initially atoms move without acceleration in the $\left|
m=0\right\rangle $ state. In the first layer, atoms split between Zeeman
sublevels $\left| m=\pm 1\right\rangle $ and start accelerating. In the
second layer atoms are partially returned to the initial state and stop
accelerating, while atoms in other sublevels (their trajectories shown by
dashed curves) leave the interferometer.\protect\smallskip\ \hspace{5in}(b)
Coupling of the atomic Zeeman sublevels by optical waves propagating along
the $y$ axis inside each layer and having frequencies and polarization
vectors $\left\{ \Omega ,{\bf \hat{z}}\right\} $ and $\left\{ \Omega +%
\protect\omega _{Z},{\bf \hat{x}}\right\} $. 
\label{f02}}
\end{minipage}
\end{figure}

In this case the reduced matrix element of the two-quantum transition is
proportional to that arising in the absence of Zeeman and hyperfine
splittings. Since alkali ground states have angular moment $J_{G}=1/2$,
selection rules allow field absorption and emission processes only where the
angular moment projection changes by at most one, and thus the two-quantum
reduced matrix elements between ground state levels vanish. As a result, a
combination of $\sigma _{+}-\sigma _{-}$ fields produce no transitions. For
this reason we assume here that the fields' frequencies and polarization
vectors are chosen as $\left\{ \Omega ,{\bf \hat{z}}\right\} ,$ $\left\{
\Omega +\omega _{Z},{\bf \hat{x}}\right\} $, while the atomic initial state
is $\left| G=1,m=0\right\rangle ,$ where $G$ is the total atomic angular
moment (see Fig. \ref{f02}b). In the atomic rest frame traveling waves
localized in the thin layer act as a pulse. For $G=1,$ a proper choice of
the pulse area allows one to split all atoms symmetrically between Zeeman
sublevels $\left| m=\pm 1\right\rangle $ (see appendix \ref{atom}) and start
their acceleration.

To stop the acceleration, one applies another set of traveling waves located
on a plane $x=+d$ to return atoms back to the $\left| m=0\right\rangle $
states. The $x$-range within which the acceleration that split the atomic
beam are active is thereby limited to the thin region $|x|<d$. One has to
distinguish the regimes of weak and strong acceleration, characterized by 
\begin{mathletters}
\label{i21}
\begin{eqnarray}
\eta &\ll &1,  \label{i21a} \\
\eta &\gg &1,  \label{i21b}
\end{eqnarray}
respectively, where 
\end{mathletters}
\begin{equation}
\eta =\Delta z/b,  \label{i22}
\end{equation}
$b$ is an atomic beam radius, and $\Delta z$ is the atom displacement during
the acceleration. When the acceleration is weak (the case shown in Fig. \ref
{f02}a), on the planes $x=-d$ and $x=+d$ the same pair of fields can be
used. Since the Zeeman splitting is equidistant, these fields drive a chain
of transitions $\left| m=\pm 1\right\rangle \rightarrow \left|
m=0\right\rangle \rightarrow \left| m=\mp 1\right\rangle ,$ and, evidently,
can not return all atoms to the $\left| m=0\right\rangle $ state.
Nevertheless, it is possible to maximize the amplitude of return to the $%
\left| m=0\right\rangle $ state. At $x>d,$ atoms that remain in the state $%
\left| m=\pm 1\right\rangle $ continue to accelerate and eventually leave
the interferometer, while atoms in the state $\left| m=0\right\rangle $ are
of further interest.

In Fig.~\ref{f03}, quadrupoles II and III act along the spatially separated
arms of the interferometer, and are adjusted such that they reverse the
transverse components of the atomic momenta. For this purpose, one may still
apply fields $\left\{ \Omega ,{\bf \hat{z}}\right\} \ $and $\left\{ \Omega
+\omega _{Z},{\bf \hat{x}}\right\} $ to transfer atoms at $x=-d$ to the
states $\left| m=\pm 1\right\rangle ,$ decelerate them between $-d<x<d,$ and
return them back to the state $\left| m=0\right\rangle $ at $x=+d$. (the
origin is assumed to be at the center of the respective quadrupole field).
Several additional arms of useless atoms in $\left| m=\pm 1\right\rangle $
states will be produced. Our calculation show that only one-eighth of the
atoms will be properly recombined to produce a grating, while seven-eighths
will be lost. To avoid the loss, we propose to use a second hyperfine
manifold, having angular momentum $G=2.$ If at $x=-d$ one applies fields $%
\left\{ \Omega ,{\bf \hat{z}}\right\} \ $and $\left\{ \Omega +\omega
_{21,10},{\bf \hat{x}}\right\} ,$ where $\omega _{21,10}$ is the frequency
of the transition $\left| G=2,m=1\right\rangle \rightarrow \left|
G=1,m=0\right\rangle ,$ then only this two-level scheme is involved, because
the frequencies $n\omega _{21,10}$ ($n\neq 1$ is integer) no longer coincide
with any atomic transition frequency. Choosing a field pulse area $\Lambda
=\pi ,$ one can transfer 100\% of the atoms at $x=-d$ into the $\left|
G=2,m=1\right\rangle $ state, accelerate the atoms, and return all of them
back to the $\left| G=1,m=0\right\rangle $ state at $x=d$ by another $\pi $%
-pulse.

After the action\ of quadrupoles I - III of Fig. \ref{f03}, the total difference of
momenta in the interferometer arms is given by 
\begin{equation}
\Delta p=\Delta p_{I}+\Delta p_{II}.  \label{i3}
\end{equation}
Momentum kicks $\Delta p_{i}$ associated with quadrupole $i$ have opposite
signs. One needs to use $\Delta p_{II}$ partially to cancel $\Delta p_{I}$.
The larger $\left| \Delta p_{II}\right| $ requires a larger field and more
severe conditions for the gradient homogeneity. As a result, for a given
desirable gratings period $\lambda _{g}$ it is better to choose 
\begin{equation}
\left| \Delta p_{I}\right| \ll \left| \Delta p_{II}\right| ,  \label{i31}
\end{equation}
such that 
\begin{equation}
\Delta p_{II}\approx \Delta p.  \label{i4}
\end{equation}
In this case the role of quadrupole I is just to split the beam into two
arms, while quadrupoles II and III are responsible almost entirely for the
grating formation.

Owing to the inhomogeneity of the field gradient, the finite time of the
interaction, and the finite angle of the atom scattering, instead of
changing the atom momentum $p$ by fixed value $\Delta p,$ one produces a
wave packet in the momentum space\ near the momentum $p+\Delta p$ with a
width that increases for larger momentum kick $\Delta p$ (or smaller $%
\lambda _{g}).$ We analyzed this effect recently for LABS produced using
resonant fields \cite{c2,c21}. The atom grating profile would be damaged if
the wave packet width becomes larger than $\hbar /b.$ In this article we
evaluate corrections to the wave function associated with the factors listed
above. For a given grating period $\lambda _{g}$, we find other important
characteristics of the problem from the requirement for corrections to be
small. These wave function corrections allow us to choose an atomic beam
aperture $b$ and velocity $u,$ the length of the interaction zone $d$, the
magnetic field gradient $B^{\prime },$ and the bias field strength $B_{s}$
to obtain a desired grating with a given accuracy.

The performance of the various Stern-Gerlach acceleration regions in the
above schemes is limited by chromatic aberration and aberrations due to
inhomogeneities of the field gradients. The detailed analysis presented in
the following sections provides a quantitative foundation to estimate these
effects, and to identify the best possible operating conditions.

\section{Atomic scattering from a linear potential with small corrections}

Our Stern-Gerlach beam splitters can be characterized by a potential

\begin{equation}
\tilde{U}\left( z,x\right) =U_{0}\left( x\right) -\tilde{f}\left( x\right) z+%
\tilde{U}_{1}\left( z,x\right),  \label{i5p}
\end{equation}

which acts on atoms propagating predominantly in the $x$-direction. The
potential acts in the narrow layer $\left| x\right| <d$, and consists of a
homogeneous part $U_{0}\left( x\right) ,$ a large linear part $-\tilde{f}%
\left( x\right) z$, and a small nonlinear addition $\tilde{U}_{1}\left(
z,x\right) .$ In the following, we analyze the propagation of matter waves
in such a potential.

A time-independent solution $\Phi \left( z,x\right) $ of the matter wave at
a given energy $E$ in the potential $\tilde{U}\left( z,x\right) $ follows
the Schr\"{o}dinger Equation 
\begin{equation}
\left[ \frac{{\bf p}^{2}}{2M}+\tilde{U}\left( z,x\right) \right] \Phi \left(
z,x\right) =E\Phi \left( z,x\right) ,  \label{i5}
\end{equation}
where ${\bf p}$ and $M$ are the momentum operator and the atomic mass. When
the potential is weak compared to the kinetic energy, which is mostly given
by the motion in $x$-direction, 
\begin{equation}
\left| U_{0}\left( x\right) \right| \ll E,  \label{i50}
\end{equation}
one can use a slowly varying amplitude approximation for the wave function.
Introducing a ''time'' $t=x/u\left( x\right) ,$ where 
\begin{equation}
u\left( x\right) =\left[ 2\left( E-U_{0}\left( x\right) \right) /M\right]
^{1/2}  \label{i501}
\end{equation}
is an atomic velocity, one can seek a solution of the form 
\begin{equation}
\Phi \left( z,x\right) =\exp \left[ i\frac{M}{\hbar }\int_{-d}^{x}dxu\left(
x\right) \right] \psi \left( z,t\right) ,  \label{i51}
\end{equation}
where $\psi \left( z,t\right) $ is the slowly varying wave function
amplitude (referred to below simply as the wave function). In the momentum
representation, $\tilde{\Psi}\left( p,t\right) $ =$\left( 2\pi \hbar \right)
^{-1/2}\int dz\exp \left( -ipz/\hbar \right) \psi \left( z,t\right) ,$ this
wave function obeys the equation 
\begin{equation}
i\hbar \partial _{t}\tilde{\Psi}=\left[ \frac{p^{2}}{2M}-i\hbar f\left(
t\right) \partial _{p}+U_{1}\left( i\hbar \partial _{p},t\right) +Q\right] 
\tilde{\Psi},  \label{i6}
\end{equation}
where $f\left( t\right) =\tilde{f}\left( ut\right) $ is a force, $%
U_{1}\left( z,t\right) =\tilde{U}_{1}\left( z,ut\right) ,\ $and $Q$
represents small terms arising from a second derivative in time and the slow
variation of the atomic velocity, 
\begin{equation}
Q=-i\frac{\hbar }{2}u^{\prime }\left( x\right) \left( 1+i%
{\displaystyle{\hbar  \over Mu^{2}}}%
\partial _{t}\right) -\frac{\hbar ^{2}}{2Mu^{2}}\partial _{t}^{2}.
\label{i7}
\end{equation}

Neglecting $U_{1}$ and $Q$ one arrives at a one dimensional Schr\"{o}dinger
equation with a time dependent spatially homogeneous force, 
\begin{equation}
i\hbar \left( \partial _{t}+f\left( t\right) \partial _{p}\right) \tilde{\Psi%
}\left( p,t\right) =\frac{p^{2}}{2M}\tilde{\Psi}\left( p,t\right)  \label{i8}
\end{equation}
Recently, this equation has been exactly solved in the coordinate
representation \cite{c8,c81,c82}. For the purposes of this article, we
derive the solution with an alternate method using the momentum
representation and an accelerated frame 
\begin{mathletters}
\label{i9}
\begin{eqnarray}
p &=&p_{0}+\delta p\left( t\right) ,  \label{i9a} \\
\delta p\left( t\right) &=&\int_{-\tau }^{t}dt_{1}f\left( t_{1}\right) ,
\label{i9b}
\end{eqnarray}
where the wave function evolves as 
\end{mathletters}
\begin{equation}
i\hbar \partial _{t}\tilde{\Psi}\left( p_{0},t\right) =\frac{\left[
p_{0}+\delta p\left( t\right) \right] ^{2}}{2M}\tilde{\Psi}\left(
p_{0},t\right) .  \label{i10}
\end{equation}
Solving this equation and returning back to the lab frame, one finds the
following common expression: 
\end{multicols}
\begin{equation}
\tilde{\Psi}\left( p,t\right) =\exp \left[ -i\int_{-\tau }^{t}dt_{1}\frac{%
\left[ p+\delta p\left( t_{1}\right) -\delta p\left( t\right) \right] ^{2}}{%
2M\hbar }\right] \tilde{\Psi}\left( p-\delta p\left( t\right) ,-\tau \right)
.  \label{i11}
\end{equation}

If Eq. (\ref{i6}) is written in the accelerated frame (\ref{i9}), the term
proportional to $p_{0}^{2}$ is responsible for the matter wave spreading.
One can neglect this term if $\tau p_{0}^{2}/2M\hbar \ll 1.$ In the case of
a diffraction-limited single-mode atomic beam, $p_{0}$ is a momentum typical
of the atomic beam spread, $p_{0}\sim \hbar /b,$ where $b$ is the radius of
the incident wave. Then, the just mentioned condition is equivalent to 
\begin{equation}
\eta _{1}=\tau /\tau _{s}\ll 1,  \label{i12}
\end{equation}
where $\tau _{s}=Mb^{2}/\hbar $ is a time characteristic of the spreading of
the matter wave. Assuming that this condition is valid we drop the quadratic
term in $p_{0}$, and arrive at the equation 
\begin{equation}
i\hbar \partial _{t}\tilde{\Psi}\left( p_{0},t\right) =\left[ \frac{\delta
p^{2}\left( t\right) }{2M}+\frac{\delta p\left( t\right) }{M}%
p_{0}+U_{1}\left( i\hbar \partial _{p_{0}},t\right) +Q\right] \tilde{\Psi}%
\left( p_{0},t\right) .  \label{i13}
\end{equation}
Seeking a solution of the form 
\begin{equation}
\tilde{\Psi}\left( p_{0},t\right) =\exp \left[ -i\phi -ip_{0}\delta z\left(
t\right) /\hbar \right] \Psi \left( p_{0},t\right) ,  \label{i14}
\end{equation}
where 
\begin{mathletters}
\label{i15}
\begin{eqnarray}
\delta z\left( t\right) &=&\int_{-\tau }^{t}dt_{1}\frac{\delta p\left(
t_{1}\right) }{M},  \label{i15a} \\
\phi &=&\int_{-\tau }^{t}dt_{1}\frac{\delta p^{2}\left( t_{1}\right) }{%
2M\hbar },  \label{i15b}
\end{eqnarray}
one finds that $\Psi \left( p_{0},t\right) $ evolves as 
\end{mathletters}
\begin{equation}
i\hbar \dot{\Psi}=\left\{ U_{1}\left[ \delta z\left( t\right) +i\hbar
\partial _{p_{0}},t\right] +Q\right\} \Psi ,  \label{i16}
\end{equation}
where 
\begin{mathletters}
\label{i17}
\begin{eqnarray}
Q &=&-i\frac{\hbar }{2}u^{\prime }\left( x\right) \left( 1+i%
{\displaystyle{\hbar  \over Mu^{2}}}%
q\right) -\frac{\hbar ^{2}}{2Mu^{2}}q^{2},  \label{i17a} \\
q &=&\exp \left[ i\phi +ip_{0}\delta z\left( t\right) /\hbar \right] \left(
\partial _{t}-f\left( t\right) \partial _{p_{0}}\right) \exp \left[ -i\phi
-ip_{0}\delta z\left( t\right) /\hbar \right] .  \label{i17b}
\end{eqnarray}

The wave function in coordinate space at the exit of the interaction zone, $%
\psi \left( z,\tau \right) ,$ is given by 
\end{mathletters}
\begin{equation}
\psi \left( z,\tau \right) =\exp \left[ -i\phi +iz\Delta p/\hbar \right]
\int \frac{dp_{0}}{\left( 2\pi \hbar \right) ^{1/2}}\exp \left[ ip_{0}\left(
z-\Delta z\right) /\hbar \right] \Psi \left( p_{0},\tau \right) ,
\label{i18}
\end{equation}
where 
\begin{equation}
\Delta p=\delta p\left( \tau \right) \text{ and }\Delta z=\delta z\left(
\tau \right)  \label{i19}
\end{equation}
is a classical change of the atomic momentum and position under the
spatially homogeneous acceleration acting for a time $2\tau .$ One can
consider an atomic grating close to sinusoidal, if the period $\lambda _{g}$
is smaller than the transverse extension $b$ of the matter wave, which means
that 
\begin{equation}
\eta _{2}=\hbar /\Delta pb\ll 1.  \label{i191}
\end{equation}

In the zeroth order approximation in $U_{1}$ and $Q,$ $\Psi
_{0}(p_{0},t)=const$ and, therefore, the zeroth-order wave function (\ref
{i18}) is given by the expression 
\begin{equation}
\psi _{0}\left( z,\tau \right) =\exp \left[ -i\phi +iz\Delta p/\hbar \right]
\psi \left( z-\Delta z,-\tau \right) ,  \label{i20}
\end{equation}
which can be obtained also from the common solution (\ref{i11}) at the
assumption (\ref{i12}). One sees that a matter wave moving with a
sufficiently large and time-independent (to neglect term $Q$) velocity $u\ $%
through a layer of the homogeneous force for a time smaller than the packet
spreading time is just displaced in the phase space along the classical
trajectory.

In the absence of the higher-order effects described below, a purely
sinusoidal grating can be formed by interfering atomic momentum components.

\section{Weak acceleration}

In the following two sections we calculate higher order effects that degrade
the ideal scattering behavior of the matter wave.

Consider an atomic beam propagating with velocity $u$ along the $x$-axis and
interacting with a quadrupole magnetic field ${\bf B}\left( z,x\right) $
produced by four currents, directed along the $y$-axis, located in the $%
\left( z,x\right) $ plane at $\left( ia,ja_{x}\right) $ $(i,j=\pm 1),$ and
given by $I_{i,j}=Iij.$ In addition to the quadrupole field, one applies a
spatially homogeneous magnetic bias field ${\bf B}_{s}={\bf \hat{z}}B_{s},$
such that the total magnetic field is given by 
\begin{mathletters}
\label{12}
\begin{eqnarray}
B_{z}\left( z,x\right) &=&-B_{0}\left\{ b_{s}+a\sum_{i,j=\pm 1}ij\left(
x-ja_{x}\right) \left/ \left[ \left( z-ia\right) ^{2}+\left( x-ja_{x}\right)
^{2}\right] \right. \right\} ,  \label{12a} \\
B_{x}\left( z,x\right) &=&B_{0}a\sum_{i,j=\pm 1}ij\left( z-ia\right) \left/ 
\left[ \left( z-ia\right) ^{2}+\left( x-ja_{x}\right) ^{2}\right] \right. ,
\label{12b}
\end{eqnarray}
where $B_{0}$ is the absolute value of the magnetic field of one current at
a distance $a\ $and $b_{s}=-B_{s}/B_{0}$. In the rest frame $\left(
x=ut\right) ,$ an atom in the internal state characterized by orbital
angular momentum $L_{G},$ electronic spin $S,$ total electronic angular
momentum $J_{G},$ nuclear spin $I,$ total angular momentum $G$, and
projection $m$ of the total angular momentum on the magnetic field direction$%
,$ moves in a potential 
\end{mathletters}
\begin{equation}
U\left( z,t\right) =\mu \left| {\bf B}\left( z,ut\right) \right| ,
\label{13}
\end{equation}
where 
\begin{eqnarray}
\mu &=&\left\{ 1+\left[ J_{G}\left( J_{G}+1\right) -L\left( L+1\right)
+S\left( S+1\right) \right] /\left[ 2J_{G}\left( J_{G}+1\right) \right]
\right\}  \nonumber \\
&&\times \left\{ \left[ G\left( G+1\right) +J_{G}\left( J_{G}+1\right)
-I\left( I+1\right) \right] /\left[ 2G\left( G+1\right) \right] \right\} \mu
_{B}m  \label{14}
\end{eqnarray}
is the projection of the total magnetic moment on the direction of ${\bf B}$%
, and $\mu _{B}$ the Bohr magneton.

We assume that the atomic beam is centered at $z=0$ and has a radius $b\ll
\min \left\{ a,a_{x}\right\} ,$ and that the $\frac{\pi }{2}$- or $\pi $%
-Raman fields, which turn the interaction with the magnetic field on and
off, are located at $x=\pm d$ $\left( d\ll \min \left\{ a,a_{x}\right\}
\right) ,$ such that the half-duration of the interaction with the potential
(\ref{13}) is $\tau =d/u.$ When $b_{s}\neq 0,$ one can expand the potential (%
\ref{13}) in the vicinity of the $\left( z=0,t=0\right) $ point. Omitting
homogeneous term of Eq.~\ref{i5p} and assuming, for simplicity, that there
is no explicit time dependence of the force, one finds 
\begin{mathletters}
\label{12}
\begin{eqnarray}
U\left( z,t\right) &=&-fz+U_{1}\left( z,t\right) ,  \label{15a} \\
U_{1}\left( z,t\right) &=&f\sum_{n=2}^{\infty }\sum_{m=0}^{\infty
}c_{nm}\left( \alpha ,b_{s}\right) \frac{z^{n}}{a^{n-1}}\left( \frac{t}{\tau
_{a}}\right) ^{2m},  \label{15b}
\end{eqnarray}
where 
\end{mathletters}
\begin{equation}
f=-\mu B^{\prime },  \label{151}
\end{equation}
and $B^{\prime }$ is the magnetic field gradient at the quadrupole center,
and $\tau _{a}=a/u,$. The dimensionless coefficients 
\begin{equation}
c_{nm}\left( \alpha ,b_{s}\right) =\frac{a^{n-1}\tau _{a}^{2m}}{fn!\left(
2m\right) !}\left. \frac{\partial ^{n+2m}U}{\partial z^{n}\partial t^{2m}}%
\right| _{z=t=0}  \label{16}
\end{equation}
depend on the quadrupole size ratio 
\begin{equation}
\alpha =a_{x}/a  \label{161}
\end{equation}
and the bias field's relative strength $b_{s}.$ For a magnetic quadrupole,
the coefficients $c_{nm},$ used in further calculations, are given by 
\begin{mathletters}
\label{162}
\begin{eqnarray}
c_{21} &=&16\alpha \left[ 16\alpha ^{2}-3b_{s}^{2}\left( \alpha
^{2}-1\right) \left( \alpha ^{2}+1\right) ^{2}\right] b_{s}^{-3}\left(
\alpha ^{2}+1\right) ^{-6},  \label{162a} \\
c_{22} &=&-16\alpha \left[ 1536\alpha ^{4}-352\alpha ^{2}b_{s}^{2}\left(
\alpha ^{2}-1\right) \left( \alpha ^{2}+1\right) ^{2}+b_{s}^{4}\left( \alpha
^{2}+1\right) ^{4}\left( 21-62\alpha ^{2}+21\alpha ^{4}\right) \right]
b_{s}^{-5}\left( \alpha ^{2}+1\right) ^{-10},  \label{162b} \\
c_{23} &=&-16\alpha \left\{ -122880\alpha ^{6}+35328b_{s}^{2}\alpha
^{4}\left( \alpha ^{2}-1\right) \left( \alpha ^{2}+1\right)
^{2}-32b_{s}^{4}\alpha ^{2}\left( \alpha ^{2}+1\right) ^{4}\left(
95-234\alpha ^{2}+95\alpha ^{4}\right) \right.  \nonumber \\
&&+\left. b_{s}^{6}\left( \alpha ^{2}+1\right) ^{6}\left[ 81\left( \alpha
^{6}-1\right) -463\alpha ^{2}\left( \alpha ^{2}-1\right) \right] \right\}
b_{s}^{-7}\left( \alpha ^{2}+1\right) ^{-14},  \label{162c} \\
c_{30} &=&2\left( \alpha ^{2}-1\right) \left( \alpha ^{2}+1\right) ^{-2}. 
\eqnum{A1d}
\end{eqnarray}
\end{mathletters}
\begin{multicols}{2}
To indicate the structure of the $c_{nm}$, in the following matrix indices $%
\left( n,m\right) $ with $c_{nm}\neq 0$ are marked by a ``V'': 
\begin{equation}
\left( 
\begin{tabular}{|c|c|c|c|c|}
\hline
& $m=0$ & 1 & 2 & 3 \\ \hline
$n=2$ &  & V & V & V \\ \hline
3 & V & V & V & V \\ \hline
4 &  & V & V & V \\ \hline
5 & V & V & V & V \\ \hline
\end{tabular}
\right) .  \label{17}
\end{equation}

We characterize the problem by dimensionless parameters 
\begin{mathletters}
\label{18}
\begin{eqnarray}
\beta &=&b/a,\,  \label{18a} \\
\delta &=&d/a,\,\,  \label{18b} \\
\varepsilon _{g} &=&\lambda _{g}/a,  \label{18c} \\
\theta &=&\lambda _{dB}/\lambda _{g},  \label{18d}
\end{eqnarray}
where 
\end{mathletters}
\begin{equation}
\lambda _{dB}=2\pi \hbar /Mu  \label{181}
\end{equation}
is the atomic de-Broglie wavelength and $\theta $ is the angle of atom
scattering. The constant force one needs to apply to achieve a given atomic
grating\ period can be found from Eqs. (\ref{i19}, \ref{i9}, \ref{i2}) to be 
\begin{equation}
f=\kappa \pi \hbar u/2d\lambda _{g},  \label{19}
\end{equation}
where the parameter $\kappa \approx 1$ for quadrupoles $II$ and $III,$ and $%
\kappa =\Delta p_{I}/\Delta p\ll 1$ for quadrupole $I$ (see Fig. \ref{f03}%
b). Consequently, the atom displacement $\Delta z=\kappa d\theta /2$ and the
parameter (\ref{i22}) is given by 
\begin{equation}
\eta =\kappa \delta \theta /2\beta ,  \label{20}
\end{equation}
while the small parameters (\ref{i12}, \ref{i191}) are given by 
\begin{mathletters}
\label{201}
\begin{eqnarray}
\eta _{1} &=&\left( 2\pi \right) ^{-1}\delta \theta \varepsilon _{g}\beta
^{-2},  \label{201a} \\
\eta _{2} &=&\varepsilon _{g}/\pi \beta .  \label{201b}
\end{eqnarray}
One can express the atomic beam and magnetic quadrupole characteristics
through the parameters (\ref{18}). For example, using Eqs. (\ref{18d}, \ref
{181}) and then (\ref{19}, \ref{151}, \ref{18b}), one finds the atom
velocity and magnetic field gradient: 
\end{mathletters}
\begin{mathletters}
\label{202}
\begin{eqnarray}
u &=&\left( 2\pi \hbar /M\lambda _{g}\right) \theta ^{-1},  \label{202a} \\
B^{\prime } &=&\left( \kappa \pi ^{2}\hbar ^{2}/\mu M\lambda _{g}^{3}\right)
\varepsilon _{g}\theta ^{-1}\delta ^{-1}.  \label{202b}
\end{eqnarray}

One can use Eq. (\ref{i16}) to calculate corrections to the unperturbed atom
wave function $\psi _{0}\left( p_{0},t\right) =\psi \left( p_{0},-\tau
\right) .$ Using the estimate 
\end{mathletters}
\begin{equation}
\hbar \partial _{p_{0}}\sim b,  \label{21}
\end{equation}
for the case of weak acceleration (\ref{i21a}), one can neglect the term $%
\delta z\left( t\right) $ in Eq. (\ref{i16}). After this, one can calculate
the first-order correction $\Psi _{1}(p_{0},\tau ),$ associated with the $%
\left( n,m\right) $ term of the expansion (\ref{i15b}). Substituting the
expression for $\Psi _{1}(p_{0},\tau )$ in Eq. (\ref{i18}) one finds the
correction in the coordinate representation 
\end{multicols}
\begin{equation}
\psi _{1}\left( z,\tau \right) =-i\left[ 2fc_{nm}d^{2m+1}\right/ \left.
\left( 2m+1\right) \hbar a^{n+2m-1}u\right] \left( z-\Delta z\right)
^{n}\psi _{0}\left( z,\tau \right) ,  \label{23}
\end{equation}
where $\psi _{0}\left( z,\tau \right) $ is given by Eq. (\ref{i20})

We now proceed to calculate the correction $\Psi _{q}(p_{0},\tau )$
associated with a term $Q$ in Eq. (\ref{i16}). In Appendix \ref
{Justification} we found the conditions under which one can neglect the
first term in the Eq. (\ref{i17a}), while the operator (\ref{i17b}) is
reduced to the expression 
\begin{equation}
q\approx -f\partial _{p_{0}}.  \label{24}
\end{equation}
Consequently, the relevant correction in coordinate space is given by 
\begin{equation}
\psi _{q}\left( z,\tau \right) =-i\left( f^{2}\tau /\hbar Mu^{2}\right)
\left( z-\Delta z\right) ^{2}\psi _{0}\left( z,\tau \right) .  \label{25}
\end{equation}
In contrast to (\ref{23}), this correction arises from the wave packet
motion through a field with a homogeneous gradient.

The parameters $\beta ,$ $\delta $ and $\theta $ have to be chosen such that
corrections (\ref{23}, \ref{25}) are small. We estimate these corrections at 
$z-\Delta z=b.$ Introducing a small parameter 
\begin{equation}
\varepsilon _{q}=\left| \psi _{q}\left( \Delta z+b,\tau \right) /\psi \left(
\Delta z+b,\tau \right) \right|  \label{251}
\end{equation}
one finds for $\theta $ 
\begin{equation}
\theta =e_{q}\varepsilon _{g}\varepsilon _{q}\delta \beta ^{-2},  \label{26}
\end{equation}
where 
\begin{equation}
e_{q}=8\pi ^{-1}\kappa ^{-2}  \label{261}
\end{equation}

Among the corrections $\psi _{1}$ for different $n$ and $m,$ the leading
terms arise from 
\begin{mathletters}
\label{262}
\begin{eqnarray}
\left( n,m\right) &=&\left( 2,m_{1}\right) ,  \label{262a} \\
\left( n,m\right) &=&\left( n_{2},0\right) ,  \label{262b}
\end{eqnarray}
as they are the first non-zero terms in the row $n=2$ or column $m=0$ of the
matrix (\ref{17}), and the other terms are higher powers of the small
parameters $\tau /\tau _{a}=\delta $ or $\left| \left( z-\Delta z\right)
/a\right| \lesssim \beta .$ Introducing the corresponding small parameters $%
\varepsilon _{1}$ and $\varepsilon _{2}$ for the relative weight of
corrections related to Eqs. (\ref{262}) and using Eq. (\ref{23}), one
arrives at equations 
\end{mathletters}
\begin{mathletters}
\label{28}
\begin{eqnarray}
\delta ^{2m_{1}}\beta ^{2} &=&e_{1}\varepsilon _{g}\varepsilon _{1},
\label{28a} \\
\beta ^{n_{2}} &=&e_{2}\varepsilon _{g}\varepsilon _{2},  \label{28b}
\end{eqnarray}
where 
\end{mathletters}
\begin{equation}
e_{1}=\left( 2m_{1}+1\right) \left( \kappa \pi \left| c_{2m_{1}}\right|
\right) ^{-1},\,\,e_{2}=\left( \kappa \pi \left| c_{n_{2}0}\right| \right)
^{-1}.  \label{29}
\end{equation}

Solving Eqs. (\ref{28}) one finds 
\begin{mathletters}
\label{30}
\begin{eqnarray}
\delta &=&\varepsilon _{1}^{1/2m_{1}}\varepsilon
_{2}^{-1/m_{1}n_{2}}\varepsilon _{g}^{\left( n_{2}-2\right)
/2m_{1}n_{2}}f_{\delta }\left( \alpha ,b_{s}\right) ,  \label{30a} \\
f_{\delta }\left( \alpha ,b_{s}\right)
&=&e_{1}^{1/2m_{1}}e_{2}^{-1/m_{1}n_{2}},  \label{30b} \\
\beta &=&\left( \varepsilon _{g}\varepsilon _{2}\right) ^{1/n_{2}}f_{\beta
}\left( \alpha ,b_{s}\right) ,  \label{30c} \\
f_{\beta }\left( \alpha ,b_{s}\right) &=&e_{2}^{1/n_{2}}  \label{30d} \\
\theta &=&\varepsilon _{1}^{1/2m_{1}}\varepsilon _{2}^{-\left(
2m_{1}+1\right) /m_{1}n_{2}}\varepsilon _{q}\varepsilon _{g}^{1+\left(
n_{2}-4m_{1}-2\right) /2m_{1}n_{2}}f_{\theta }\left( \alpha ,b_{s}\right) ,
\label{30e} \\
f_{\theta }\left( \alpha ,b_{s}\right) &=&e_{1}^{1/2m_{1}}e_{2}^{-\left(
2m_{1}+1\right) /m_{1}n_{2}}e_{q}.  \label{30f}
\end{eqnarray}
\end{mathletters}
\begin{multicols}{2}
One can use Eqs. (\ref{30}) to estimate the efficiency of the beam splitter
with arbitrary non-linearity.

\begin{figure}
\begin{minipage}{0.99\linewidth}
\begin{center}
\epsfxsize=.95\linewidth \epsfbox{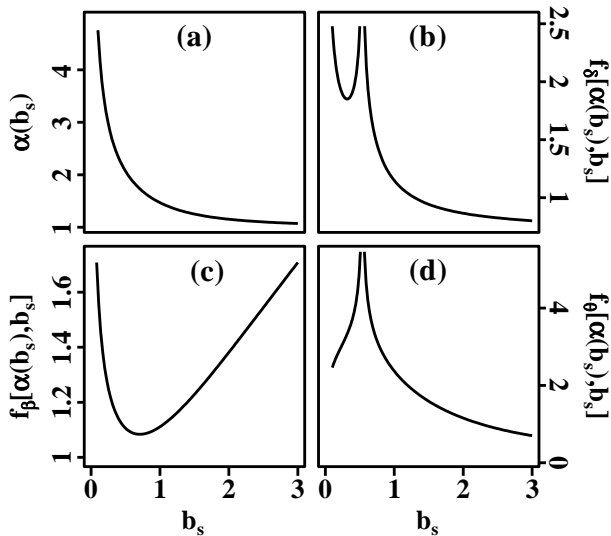}
\end{center}
\end{minipage}
\begin{minipage}{0.99\linewidth} \caption{Dependence of the quadrupole size ratio $\left( a\right) ,$ width
of the acceleration zone $\left( b\right) ,$ atomic beam radius $\left(
c\right) ,$ and scattering angle $\left( d\right) $ on the relative strength 
$b_{s}$ of bias magnetic field. 
\label{f04}}
\end{minipage}
\end{figure}

We now return to the quadrupole configuration at hand. The non-linearities
depend on the ratio of the quadrupole sizes $\alpha $ and the relative
strength of the bias magnetic field $b_{s}.$ These quantities can be used to
diminish the role of the non-linearities. One can choose \ them such that
either the coefficient $c_{21}$ or the coefficient $c_{30}$ vanishes. Our
calculations show that it is more effective to choose $c_{21}=0$; therefore,
in the remainder of this article, we consider that case. Equation 
\begin{equation}
c_{21}=0  \label{32}
\end{equation}
determines the ratio of the size $\alpha $ as a function of $b_{s}.$ The
function $\left. \alpha \left( b_{s}\right) \right| _{c_{21}=0}$ is shown in
Fig. \ref{f04}a.

For $c_{21}=0$ the lowest-order correction is the term corresponding to $%
m_{1}=2,$ $n_{2}=3$ (see matrix (\ref{17})). The functions $f\left[ \alpha
\left( b_{s}\right) ,b_{s}\right] $ are plotted in Fig. \ref{f04} for $%
\kappa =1.$ These functions have notable features for the relative field
strength $b_{s}\approx 0.537,$ where the ratio of quadrupole size (\ref{161}%
) is $\alpha \approx 1.99.$ These singular features arise because at $%
b_{s}\approx 0.537$ - in addition to condition (\ref{32}) - the
non-linearity that is quadratic in space and 4th-order in time also
vanishes, i. e. 
\begin{equation}
c_{22}=0.  \label{34}
\end{equation}

So, one has to consider the next term, $\left( 2,3\right) ,$ in the row $n=2$
of the Table (\ref{17}). When $m_{1}=n_{2}=3$ and $\left( \alpha
,b_{s}\right) $ is a root of Eqs. (\ref{32}, \ref{34}), one finds from Eqs. (%
\ref{30}): 
\begin{mathletters}
\label{35}
\begin{eqnarray}
\delta &=&1.45\varepsilon _{1}^{1/6}\varepsilon _{2}^{-1/9}\varepsilon
_{g}^{1/18},  \label{35a} \\
\beta &=&0.871\left( \varepsilon _{2}\varepsilon _{g}\right) ^{1/3},
\label{35b} \\
\theta &=&3.19\kappa ^{-25/18}\varepsilon _{1}^{1/6}\varepsilon
_{2}^{-7/9}\varepsilon _{q}\varepsilon _{g}^{7/18}.  \label{35c}
\end{eqnarray}

\section{Strong acceleration.}

When $\eta \gg 1,$ one can expand the nonlinear part of the potential $U_{1}$
in Eq. (\ref{i16}) in the operator $i\hbar \partial _{p_{0}}.$ The
zeroth-order term $U_{1}\left[ \delta z\left( t\right) ,t\right] $ depends
only on time and, therefore, changes the phase of the atomic wave function (%
\ref{i14}) to the value 
\end{mathletters}
\begin{equation}
\phi =\frac{1}{\hbar }\int_{-\tau }^{t}dt_{1}\left[ \frac{\delta p^{2}\left(
t_{1}\right) }{2M}+U_{1}\left[ \delta z\left( t_{1}\right) ,t_{1}\right] %
\right] .  \label{36}
\end{equation}
After a phase transformation one arrives at the equation

\begin{equation}
i\hbar \dot{\Psi}=\left\{ U_{z}\left( t\right) i\hbar \partial
_{p_{0}}+Q\right\} \Psi ,  \label{37}
\end{equation}
where 
\begin{equation}
U_{z}\left( t\right) =\left. \frac{\partial U_{1}\left( z,t\right) }{%
\partial z}\right| _{z=\delta z\left( t\right) }.  \label{38}
\end{equation}
Since one can still neglect the first term in Eq. (\ref{i17a}) and use Eq. (%
\ref{24}) for the operator $q$ (see Appendix \ref{Justification}), the
previously calculated correction $\psi _{q}\left( z,\tau \right) ,$
associated with the $Q$-term, and, therefore, Eq. (\ref{26}) are still
valid. Using the expansion (\ref{15b}) one obtains a series for the operator
(\ref{38}). Keeping only the $\left( n,m\right) $ term of this series in the
right-hand-side of Eq. (\ref{37}), one finds that the corresponding
correction in the coordinate representation is given by 
\end{multicols}
\begin{equation}
\psi _{1}\left( z,t\right) =-i\left[ nc_{nm}I_{nm}f^{n}d^{2\left( m+n\right)
-1}\left( z-\Delta z\right) /\hbar a^{2m+n-1}u^{2n-1}\left( 2M\right) ^{n-1}%
\right] \psi _{0}\left( z,t\right) ,  \label{39}
\end{equation}
where 
\begin{equation}
I_{nm}=\int_{-1}^{1}d\xi \xi ^{2m}\left( 1+\xi \right) ^{2\left( n-1\right)
}.  \label{40}
\end{equation}
Requiring the magnitude of this correction at $z=\Delta z+b$ to be $%
\varepsilon _{1}$- and $\varepsilon _{2}$-times smaller than the
zeroth-order solution $\psi _{0}\left( z,t\right) $ for $\left( n,m\right)
=\left( 2,m_{1}\right) $ and $\left( n_{2},0\right) ,$ respectively, and
expressing the atom velocity $u$ and the force $f$ through parameters (\ref
{18}), one obtains equations 
\begin{mathletters}
\label{41}
\begin{eqnarray}
\beta \delta ^{1+2m_{1}}\theta &=&e_{1}\varepsilon _{g}\varepsilon _{1},
\label{41a} \\
\beta \left( \delta \theta \right) ^{n_{2}-1} &=&e_{2}\varepsilon
_{g}\varepsilon _{2},  \label{41b}
\end{eqnarray}
where for this case we define parameters $e_{i}$ as 
\end{mathletters}
\begin{equation}
e_{1}=8/\kappa ^{2}\pi c_{2m_{1}}I_{2m_{1}},\,\,e_{2}=2^{3n_{2}-2}/\kappa
^{n_{2}}\pi n_{2}c_{n_{2}0}I_{n_{2}0}.  \label{42}
\end{equation}
Equations (\ref{26}, \ref{41}) constitute a system for three variables $%
\delta ,$ $\beta ,$ $\theta ,$ which has the solution 
\begin{mathletters}
\label{43}
\begin{eqnarray}
\delta &=&\varepsilon _{1}^{\left( 2n_{2}-3\right) /2\gamma }\varepsilon
_{2}^{-1/2\gamma }\left( \varepsilon _{g}/\varepsilon _{q}\right) ^{\left(
n_{2}-2\right) /2\gamma }f_{\delta }\left( \alpha ,b_{s}\right) ,
\label{43a} \\
f_{\delta }\left( \alpha ,b_{s}\right) &=&e_{1}^{\left( 2n_{2}-3\right)
/2\gamma }e_{2}^{-1/2\gamma }e_{q}^{-\left( n_{2}-2\right) /2\gamma },
\label{43b} \\
\beta &=&\varepsilon _{1}^{\left( n_{2}-1\right) /\gamma }\varepsilon
_{2}^{-\left( m_{1}+1\right) /\gamma }\varepsilon _{g}^{\left(
n_{2}-2\right) \left( m_{1}+1\right) /\gamma }\varepsilon _{q}^{m_{1}\left(
n_{2}-1\right) /\gamma }f_{\beta }\left( \alpha ,b_{s}\right) ,  \label{43c}
\\
f_{\beta }\left( \alpha ,b_{s}\right) &=&e_{1}^{\left( n_{2}-1\right)
/\gamma }e_{2}^{-\left( m_{1}+1\right) /\gamma }e_{q}^{m_{1}\left(
n_{2}-1\right) /\gamma },  \label{43d} \\
\theta &=&\varepsilon _{1}^{-\left( 2n_{2}-1\right) /2\gamma }\varepsilon
_{2}^{\left( 4m_{1}+3\right) /2\gamma }\left( \varepsilon _{g}/\varepsilon
_{q}\right) ^{\left[ 2\left( m_{1}+1\right) -n_{2}\right] /2\gamma
}f_{\theta }\left( \alpha ,b_{s}\right) ,  \label{43e} \\
f_{\theta }\left( \alpha ,b_{s}\right) &=&e_{1}^{-\left( 2n_{2}-1\right)
/2\gamma }e_{2}^{\left( 4m_{1}+3\right) /2\gamma }e_{q}^{-\left[ 2\left(
m_{1}+1\right) -n_{2}\right] /2\gamma }  \label{43f}
\end{eqnarray}
where $e_{q}$ is defined by Eq. (\ref{261}) and $\gamma =n_{2}-2+m_{1}\left(
2n_{2}-3\right) $.
\end{mathletters}

\begin{figure}
\begin{minipage}{0.99\linewidth}
\begin{center}
\epsfxsize=.95\linewidth \epsfbox{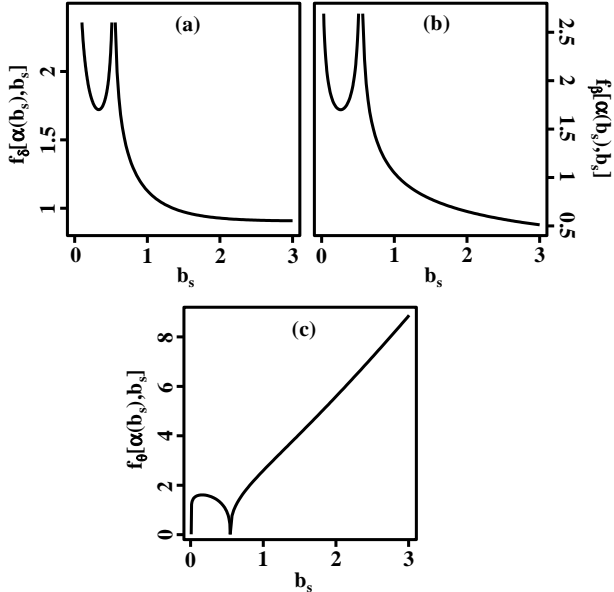}
\end{center}
\end{minipage}
\begin{minipage}{0.99\linewidth} \caption{Strong acceleration regime. Dependences of the width of the
acceleration zone $\left( a\right) ,$ atomic beam radius $\left( b\right) ,$
and scattering angle $\left( c\right) $ on the bias magnetic field relative
strength $b_{s}$. 
\label{f05}}
\end{minipage}
\end{figure}

\begin{multicols}{2}
The further consideration is the same as in the previous Section. If one
chooses the\ quadrupole axes ratio $\alpha $ such that the potential's
quadratic term in space and time vanishes, i.e. if $\alpha \equiv \alpha
\left( b_{s}\right) ,$ where the function $\alpha \left( b_{s}\right) $ is
defined explicitly by Eq. (\ref{32}) and shown in Fig. \ref{f04}a, then
again $m_{1}=2$ and $n_{2}=3.$ The functions $f\left[ \alpha \left(
b_{s}\right) ,b_{s}\right] $ are shown in Fig. \ref{f05}. 
At the point of divergence, $b_{s}\approx 0.537,$ choosing $m_{1}=n_{2}=3,$
one finds 
\begin{mathletters}
\label{45}
\begin{eqnarray}
\delta &=&1.44\kappa ^{-1/20}\varepsilon _{1}^{3/20}\left( \varepsilon
_{g}/\varepsilon _{2}\varepsilon _{q}\right) ^{1/20},  \label{45a} \\
\beta &=&1.47\kappa ^{-2/5}\varepsilon _{1}^{1/5}\left( \varepsilon
_{g}/\varepsilon _{2}\right) ^{2/5}\varepsilon _{q}^{3/5},  \label{45b} \\
\theta &=&1.70049\kappa ^{-5/4}\varepsilon _{2}^{3/4}\left( \varepsilon
_{g}/\varepsilon _{1}\varepsilon _{q}\right) ^{1/4}.  \label{45c}
\end{eqnarray}
\end{mathletters}

\section{Discussion}

The scattering of an atomic center-of-mass motion wave packet from a narrow
layer of quadrupole and bias magnetic fields is analyzed. The combination of
these fields produces an approximately linear potential for atoms. It was
shown that for a purely linear potential, infinitely small atomic de-Broglie
wave length, and time of interaction smaller than the wave packet spreading
time, the wave packet scatters along the classical trajectory, changing its
momentum by a given amount $\Delta p$ {\em without any wave packet
deformation.} Only when this regime of scattering is realized, at least
approximately, one can expect that an interference between scattered and
recombined components of the atomic wave function leads to a sinusoidal
atomic grating of nanometer-scale periodicity.

In this paper we calculated corrections to the atomic wave function caused
by potential non-linearities and a small atomic de-Broglie wave length. When
the grating period $\lambda _{g}$, the quadrupole size $a$ along the grating
formation direction, the relative strength of the bias magnetic field $%
b_{s}, $ and the relative weight\ of the corrections\ caused by
non-linearities $\left( \varepsilon _{1}\text{ and }\varepsilon _{2}\right) $
and by a finite de-Broglie wave length $\left( \varepsilon _{q}\right) $ are
given, one can use our analysis to determine the atomic beam velocity $u$
and transverse size $b,$ the magnetic field gradient $B^{\prime }$and the
bias field strength $B_{s},$ the thickness of the interaction layer $d,$ and
the ratio of the quadrupole axes $\alpha $ that will minimize nonlinear
effects.

One can consider nonlinear corrections to the wave function as a spherical
aberration of the beam splitter. There are two more types of aberrations,
namely chromatic aberration and the atomic beam angular divergence.
Chromatic aberration arises from averaging the grating over the atomic
longitudinal velocity $u.$ Since the atom momentum change $\Delta p$ is
proportional to the time of acceleration $\tau \propto 1/u$, the grating
period [by Eq. (\ref{i2})] is linear in $u$,

\begin{equation}
\lambda _{g}\propto u.  \label{451}
\end{equation}
To achieve a grating with a given period, one has to use a monovelocity
beam. Chromatic aberration occurs as a consequence of a small but finite
width of the velocity distribution. To be specific, consider a Gaussian
distribution 
\begin{equation}
W\left( u\right) =\pi ^{-1/2}\left( \sigma _{0}\right) ^{-1}\exp \left[
-\left( \sigma /\sigma _{0}\right) ^{2}\right] ,  \label{46}
\end{equation}
where $u_{0}$ is a mean velocity, $\sigma =\left( u-u_{0}\right) /u_{0},$
and, for a beam having longitudinal temperature $\Theta ,$ $\sigma
_{0}=u_{0}^{-1}\left( 2k_{B}\Theta /M\right) ^{1/2}$ is the small relative
width of the distribution. For a given velocity, atom interference results
in a term in the atom density, 
\begin{equation}
\rho \propto \cos \left( \phi _{tot}\right) ,  \label{461}
\end{equation}
where $\phi _{tot}$ is a total phase difference of the wave functions in two
arms of an interferometer. The sensitivity of the grating to the velocity
results from the velocity dependence of $\phi _{tot}$. Our purpose was to
create phase difference 
\begin{equation}
\phi _{z}=k_{g}z\approx \phi _{z0}\left( 1-\sigma +\sigma ^{2}\right) ,
\label{47}
\end{equation}
where $\phi _{z0}$ is the phase at $u=u_{0},$ $k_{g}=2\pi /\lambda _{g}$ is
a wave number associated with the grating period, and we take into account
Eq. (\ref{451}).

The largest phase that the atoms acquire is the Talbot phase associated with
the atomic kinetic energy. This phase leads to Talbot oscillations \cite
{mTlbt} of the interference pattern, first observed in a Na beam \cite{oTlbt}%
. Averaging over the longitudinal velocity is equivalent to the averaging
over the Talbot phase. One has to choose an interferometric scheme, in which
Talbot phases can be compensated. Evidently, the symmetric configuration of
the interferometer satisfies this requirement. Moreover, for this
configuration one compensates not only Talbot phases acquired during the
free particles propagation but also those associated with the atoms'
acceleration inside the beam splitters, independently of the acceleration
time.

The next contribution to $\phi _{tot}$, which we denote as $\phi _{U},$ is
caused by the fact that accelerated atom has slightly different velocity $%
u\left( x\right) $ during acceleration, owing to the homogeneous part of the
potential $U_{0}\left( x\right) $ [see phase factor in Eq. (\ref{i51})].
Since, during acceleration in the symmetric configuration, atoms in two arms
are in substates having opposite magnetic quantum numbers, they acquire
phases of the same magnitude and opposite sign. Therefore, the grating phase
is twice as large as the phase along a given arm. For the arm $1,$ expanding
Eq. (\ref{i501}) to first order in $U_{0}\left( x\right) /E,$ one finds 
\begin{equation}
\phi _{U}=2\int_{-d}^{d}\frac{dx}{\hbar u}U_{0}\left( x\right) .  \label{48}
\end{equation}
This phase behaves as $1/u,$ i.e., $\phi _{U}\approx \phi _{U0}\left(
1-\sigma +\sigma ^{2}\right) .$ For the potential produced by the magnetic
quadrupole, $U_{0}\left( x\right) \approx fab_{s}\left( 1+\alpha ^{2}\right)
^{2}/8\alpha ,$ and therefore 
\begin{equation}
\phi _{U0}=\pi b_{s}\left( 1+\alpha ^{2}\right) ^{2}/4\alpha \varepsilon
_{g}.  \label{49}
\end{equation}
In the case of weak acceleration, there are no other contributions to the
phase associated with chromatic aberration. Owing to the large value of this
phase, even for small widths of the velocity distribution, the grating can
be washed out after averaging over velocities. To avoid this situation, we
propose to insert one more element into the interferometer, a region of
homogeneous magnetic field. In this region an atomic wave function acquires
a phase $\phi _{c}\propto u^{-1},$ i.e., $\phi _{c}=\phi _{c0}\left(
1-\sigma +\sigma ^{2}\right) .$ Choosing this phase to compensate the phase $%
\phi _{U}$ (\ref{48}), one finds for the grating averaged over velocities 
\end{multicols}
\begin{equation}
\bar{\rho}\propto \left( 1+\zeta ^{2}\right) ^{-1/4}\exp \left[ -\left(
kz\sigma _{0}\right) ^{2}/4\left( 1+\zeta ^{2}\right) \right] \cos \left\{
kz-0.5\tan ^{-1}\left( \zeta \right) +\left( kz\sigma _{0}\right) ^{2}\zeta
/4\left( 1+\zeta ^{2}\right) \right\} ,  \label{52}
\end{equation}
where $\zeta =kz\sigma _{0}^{2}.$ When 
\begin{equation}
\zeta \ll 1,  \label{53}
\end{equation}
we see that averaging over velocity creates a Gaussian envelope of the
grating profile having a half width 
\begin{equation}
s=\pi ^{-1}u\lambda _{g}\left( M\ln 2/2k_{B}\Theta \right) ^{1/2},
\label{541}
\end{equation}
which is an inhomogeneous coherence half-length. The small value of $s$ is
the main problem of the technique we consider here.

The situation becomes more complicated for the strong acceleration regime,
where one has to include the phase caused by the non-linear part of the
potential [see the second term in the brackets of Eq. (\ref{36})]. The
contribution to this phase arising from the $\left( n,m\right) $ term of the
potential expansion (\ref{15b}) is given by 
\begin{mathletters}
\label{542}
\begin{eqnarray}
\phi &=&\left( u_{0}/u\right) ^{2n+1}\phi _{0},  \label{542a} \\
\phi _{0} &=&2^{1-3n}\pi \kappa ^{n+1}c_{nm}\left( \alpha ,b_{s}\right)
I_{2n,2m}\theta ^{n}\delta ^{2m+n}\varepsilon _{g}^{-1}.  \label{542b}
\end{eqnarray}
Leading terms here are those associated with $\left( n,m\right) $ given by
Eq. (\ref{262})$,$ for which we denote phases as $\phi _{10}$ and $\phi
_{20}.$ Including only these phases, one finds that the total phase is given
by 
\end{mathletters}
\begin{equation}
\phi _{tot}=\left( \phi _{z0}+\phi _{U0}+\phi _{c0}\right) \left(
u_{0}/u\right) +\phi _{10}\left( u_{0}/u\right) ^{5}+\phi _{20}\left(
u_{0}/u\right) ^{2n_{2}+1}.  \label{543}
\end{equation}
\begin{multicols}{2}
Since new terms are not proportional to $u^{-1},$ one cannot choose a
compensating phase $\phi _{c0}$ to reduce $\phi _{tot}$ only to the
desirable phase (\ref{47}), but one can choose $\phi _{c0}$ to offset the
most dangerous contribution to the aberration, that linear in $\sigma $.
Cancellation of this term occurs when 
\begin{equation}
\phi _{c0}=-\phi _{U0}-5\phi _{10}-\left( 2n_{2}+1\right) \phi _{20}.
\label{544}
\end{equation}
For this choice one recovers expression (\ref{52}) for the grating profile,
in which one has to insert the phase shift $-4\phi _{10}-2n_{2}\phi _{20},$
and change the parameter $\zeta $ to the value $\zeta =\left( k_{g}z+10\phi
_{10}+n_{2}\left( 2n_{2}+1\right) \phi _{20}\right) \sigma _{0}^{2}.$

We next consider the role of the atomic beam's angular divergence. If the
angle between the initial momentum and the $\left( x,y\right) $ plane is
non-zero, then the atom enters the acceleration zone at a non-zero momentum
projection along $z,$ $p_{in}=Mu\theta _{b},$ where $\theta _{b}$ is of the
order of the angular divergence. For the weak scattering regime at $%
p_{in}\neq 0,$ one has to shift momentum change in the definition of phase (%
\ref{i15b}) as $\delta p\left( t\right) \rightarrow p_{in}+\delta p\left(
t\right) .$ Phases quadratic in $p_{in}$ are the same for both arms of
interferometer, while the $p_{in}$-independent part is analyzed above.
Therefore, we can consider only the parts linear in $p_{in}$, $\phi _{Di},$
for which, using Eqs. (\ref{i15}), one finds $\phi _{Di}=p_{in}z_{i}\left(
t\right) /\hbar ,$ where $z_{i}\left( t\right) $ is the atomic $z$%
-coordinate along arm $i$ for $p_{in}=0$\ ($i=1$ or $2,$ see Fig. \ref{f03}%
b). Evidently $\phi _{Di}$ is a Doppler phase. The Doppler phase difference, 
\begin{equation}
\phi _{D}=2\pi \kappa \theta _{b}\left( x_{e}-x\right) /\lambda _{g},
\label{55}
\end{equation}
vanishes at the echo point, $x_{e}\approx L\left( 1+f_{I}/f\right) $, where $%
L$ is the distance between quadrupoles along $x$ axis, $f_{I}\ll f$ is a
force in quadrupole $I$. A cancellation of the Doppler phase at the
interference plane is a common property of an atom interferometer \cite
{cohen}. For the quadrupole beam-splitter, we prove that cancellation occurs
for a finite interaction time, while for optical beam splitters, involving
couterpropagating waves, Eq. (\ref{55}) is valid only in the Raman-Nath
approximation. Owing to the Doppler phase cancellation, the only requirement
for the weak acceleration regime is that the angular divergence be less than
the scattering angle, 
\begin{equation}
\theta _{b}<\left( f_{I}/f\right) \theta .  \label{551}
\end{equation}

The situation changes for the strong acceleration regime, again owing to the
phase (\ref{36}) sensitivity to the non-linear part of potential. For $%
p_{in}\neq 0,$ in the quadrupole $II,$ $\delta z\left( t\right)
=M^{-1}\left( p_{in}\left( t+\tau \right) -f\left( t+\tau \right)
^{2}/2\right) .$ Assuming that the change of the atomic position is small,
one finds that the additional Doppler phase, which is linear in $p_{in}$ and
associated with the $\left( n,m\right) $ term in the potential expansion (%
\ref{15b}), is given by 
\begin{mathletters}
\label{552}
\begin{eqnarray}
\delta \phi _{D} &=&\theta _{b}/\theta _{nm},  \label{552a} \\
\theta _{nm} &=&2^{3n-2}\kappa ^{-n}\left( n\pi c_{nm}I_{2n-1,2m}\right)
^{-1}\varepsilon _{g}\delta ^{-2m-n}\theta ^{1-n}.  \label{552b}
\end{eqnarray}
Requiring this phase to be small for leading terms in the series (\ref{15b}%
), one obtains a condition for the angular divergence, 
\end{mathletters}
\begin{equation}
\theta <\min \left\{ \left( f_{I}/f\right) \theta ,\theta _{1},\theta
_{2}\right\} ,  \label{553}
\end{equation}
where $\theta _{1}=\theta _{2,m_{1}},\theta _{2}=\theta _{n_{2},0}$.

One can use expression (\ref{55}) to estimate the thickness $\Delta x$ of
the layer along the $x$-axis where interference occurs. Requiring $\phi
_{D}\lesssim 1,$ one finds 
\begin{equation}
\Delta x\lesssim \lambda _{g}/2\pi \kappa \theta _{b}.  \label{56}
\end{equation}

We can also estimate the width of the layers $\delta x,$ in which one
excites atoms to start and to stop the acceleration. When this width is
non-zero, the time of acceleration is not fixed, and the momentum change $%
\Delta p$ is spread across a range of width $f\delta x/u$. This width should
be smaller than $\hbar /s,$ i.e. 
\begin{equation}
\delta x\lesssim \overline{\delta x}=2d\lambda _{g}/\kappa \pi s.  \label{57}
\end{equation}
Knowing $\overline{\delta x}$ one can estimate parameter (\ref{0}) as $%
\omega _{Z}\tau _{i}=\phi _{U0}\overline{\delta x}/2d.$

As an example for the scheme described in this paper, consider a beam of $^{%
\text{87}}$Rb atoms $\left( L=0,\,\,S=1/2,\,\,I=3/2\right) ,$ initially
pumped into the state $\left| G=1,m=0\right\rangle $ and having a
longitudinal temperature $\Theta =1\mu K.$ As we explained in the
Introduction, acceleration occurs near the centers of the quadrupoles.
Quadrupole $I$ just splits an atom trajectory into two arms (see Fig. \ref
{f03}b). The main part of the atomic momentum change $\Delta p$ is acquired
in quadrupoles $II$ and $III.$ We present results of calculations for the
last quadrupoles, where one can expect the most severe restrictions for the
system parameters. Inside \ the quadrupoles, the atom is accelerated in the
states $\left| G=2,m=\pm 1\right\rangle $ with a magnetic moment $\mu =\pm
\mu _{B}/2.$

It is not evident in advance what role the different types of nonlinear
corrections to the atomic wave function play. This role depends on the
acceleration regime, bias field strength, and the quadrupole geometry. One
notices, nevertheless, that all parameters of the system depend on three
variables. Instead of $\varepsilon _{1},$ $\varepsilon _{2},$ $\varepsilon
_{q},$ one can choose any other three linearly independent parameters. It is
reasonable to choose variables which are most severely restricted, and
consider how large they can be such that nonlinear corrections are still
small.

For a weak acceleration regime we choose the coherence length $s,$ the
layers' thickness $\overline{\delta x}$ and the ratio $\eta $ of the atom
displacement and the beam radius as independent variables, given by Eqs. (%
\ref{20}, \ref{541}, \ref{57}), respectively. We found that for $\eta =0.1$
and quadrupole size $a=1\,$cm, nonlinear corrections do not rise above $10\%$
if $s=40\mu m$ and $\overline{\delta x}=15\mu m.$ The beam and fields
parameters corresponding to this choice are given in Table \ref{t1}.
\end{multicols}

\begin{center}
\begin{table}[tbp]
\caption{Parameters of the beam of $^{\text{87}}$Rb atoms and the magnetic
quadrupole \ that one can choose to obtain a grating of $\protect\lambda %
_{g}=100\,$nm period (case 1) and $10$nm (cases 2 and 3): $b$\ is the
half-width of the incident wave packet; $d$\ is the half-thickness of the
region in which acceleration occurs; $u$\ is the beam velocity; $B^{\prime }$%
\ is magnetic field gradient; $b_{s}$\ and $B_{s}$\ are the relative and
absolute strengths of the bias magnetic field; $\protect\alpha $\ is the
aspect ratio of the quadrupole size; $s$\ is the coherence half-length; $%
\protect\varepsilon _{1},$\ $\protect\varepsilon _{2},$\ and $\protect%
\varepsilon _{q}$\ are relative weights of corrections to the atomic wave
function; $\protect\theta $\ is the scattering angle; $\protect\theta _{1}$\
and $\protect\theta _{2}$\ are upper bounds for the beam angular divergence
arising in the strong acceleration regime; $\protect\phi _{U0},$\ $\protect%
\phi _{10},$\ and $\protect\phi _{20}$\ are atom grating phases caused by
the homogenous and nonlinear parts of the potential; $\overline{\protect%
\delta x}$\ is the thickness of the region in which one starts and stops
atomic acceleration; $\protect\omega _{Z}\protect\tau _{i}$\ is the
parameter (\ref{0}); $P_{0}$\ and $P_{0}^{\prime }$\ are the geometric
averages of traveling wave powers one should apply to split atoms between $%
m=\pm 1$\ Zeeman sublevels and to produce a $\protect\pi $-pulse on the
transition $\left| G=1,m=0\right\rangle \rightarrow \left| G=2,m=\pm
1\right\rangle ,$ respectively [$P_{0}$\ and $P_{0}^{\prime }$ are evaluated
using Eqs. (\ref{aa7}, \ref{aa12}) where we put $\protect\delta z=0.1$ cm
and $\Delta _{J_{H}1}^{\left( 2\right) }=2\protect\pi \times 1GHz$]; $%
\protect\eta $\ is the ratio of the atom displacement during acceleration
and the beam radius, $\protect\eta _{1}\ $is the ratio of the time of
acceleration and the time of wave packet spreading, $\protect\eta _{2}\ $is
the ratio of the momentum distribution width and the momentum change during
acceleration; parameters $\protect\eta _{3},$\ $\protect\eta _{4},$\ $%
\protect\eta _{5}$ verify the validity of the slowly varying amplitude
approximation (see Appendix \ref{Justification}). The parameters $\protect%
\theta _{1},$ $\protect\theta _{2},$ $\protect\phi _{10},$ $\protect\phi %
_{20},$ $\protect\eta _{3},$ $\protect\eta _{4}$ are not relevant in the
weak acceleration regime. Parameters marked with stars in the Table are
chosen as independent. All other parameters depend on these.}
\label{t1}
\end{table}
\begin{tabular}{|c|c|c|c|}
\hline
case \# & 1\label{m1=2n2=3SXetta.nb} & 2\label{m1=2n2=3sSXB.nb} & 3\label%
{m1=n2=3SXB.nb} \\ \hline
regime & weak acceleration & strong acceleration & strong acceleration \\ 
\hline
$\lambda _{g}\left[ \text{nm}\right] $ & 100 & 10 & 10 \\ \hline
$b[\mu m]$ & 104 & 3.5$^{\text{*}}$ & 4$^{\text{*}}$ \\ \hline
$d\left[ \text{cm}\right] $ & 0.94 & 0.55 & 0.88 \\ \hline
$u\left[ \text{m/s}\right] $ & 21 & 18 & 21 \\ \hline
$B^{\prime }\left[ \text{Gs/cm}\right] $ & 79 & 1190 & 847 \\ \hline
$b_{s}$ & 0.1 & 1 & 0.54 \\ \hline
$B_{s}\left[ \text{Gs}\right] $ & 116 & 1010 & 705 \\ \hline
$\alpha $ & 4.8 & 1.5 & 1.99 \\ \hline
$s\left[ \mu m\right] $ & 40$^{\text{*}}$ & 3.5$^{\text{*}}$ & 4$^{\text{*}}$
\\ \hline
$\varepsilon _{1}$ & 0.087 & 0.088 & 0.11 \\ \hline
$\varepsilon _{2}$ & 0.027 & 0.0073 & 0.017 \\ \hline
$\varepsilon _{q}$ & 0.0098 & 0.0022 & 0.0016 \\ \hline
$\theta $ & 2.2$\times 10^{-3}$ & 0.025 & 0.021 \\ \hline
max$\left\{ \theta _{1},\theta _{2}\right\} $ &  & 0.015 & 0.0045 \\ \hline
$\phi _{U0}$ & 9.2$\times 10^{5}$ & 5.3$\times 10^{6}$ & 5.2$\times 10^{6}$
\\ \hline
$\phi _{10}$ &  & 6.8 & 11 \\ \hline
$\phi _{20}$ &  & 1.4 & 4.0 \\ \hline
$\overline{\delta x}\left[ \mu m\right] $ & 15$^{\text{*}}$ & 10$^{\text{*}}$
& 14$^{\text{*}}$ \\ \hline
$\omega _{Z}\tau _{i}$ & 733 & 4800 & 4200 \\ \hline
$P_{0}\left[ mW\right] $ & 0.15 & 0.13 & 0.15 \\ \hline
$P_{0}^{\prime }\left[ mW\right] $ & 0.12 & 0.11 & 0.12 \\ \hline
$\eta $ & 0.1$^{\text{*}}$ & 20 & 24 \\ \hline
$\eta _{1}$ & 1.9$\times 10^{-4}$ & 0.11 & 0.12 \\ \hline
$\eta _{2}$ & 3.1$\times 10^{-4}$ & 9.1$\times 10^{-4}$ & 8.0$\times 10^{-4}$
\\ \hline
$\eta _{3}$ &  & 0.050 & 0.43 \\ \hline
$\eta _{4}$ &  & 0.0075 & 0.018 \\ \hline
$\eta _{5}$ & 0.18 & 12 & 37 \\ \hline
\end{tabular}
\vspace{-0.2in}
\end{center}
\begin{multicols}{2}
In the strong acceleration regime it makes sense to choose a grating target
period $\lambda _{g}=10$ nm. As independent variables, we choose the
coherence half-length $s,$ the beam radius $b,$ and the layer thickness $%
\overline{\delta x}.$ To achieve large values of the parameter $\eta $ and
yet small weights for the corrections we choose $s=b=3.5\mu m$ and $%
\overline{\delta x}=10\mu m.$ Data for this case are given in the second
column of Table \ref{t1}. A slightly better situation arises if one chooses
conditions, in which nonlinearities quadratic in space, second order and
fourth order in time vanish. These conditions arise for $\alpha \approx 1.99$
and $b_{s}\approx 0.54$, which are roots of Eqs. (\ref{32}, \ref{34}). Data
for this case are given in the third column of Table \ref{t1}.

Beam splitters can operate also in a pulsed regime. If the whole layer 
\begin{equation}
\left| x\right| \lesssim d  \label{58}
\end{equation}
is illuminated by a short raman pulse, atoms are split between $m\not=0$
Zeeman sublevels and start to accelerate. A second, time-delayed pulse stops
the acceleration, and produces two groups of states in the $m=0$ sublevel
with different momenta. When these groups recombine at the interference
plane, a pulsed atom grating is generated. The pulsed grating can be
repeated with some repetition rate. The pulse regime will have restricted
application in lithography, because in the time intervals between the pulses
the flow of atoms will continue producing a uniform background. However,
this flow could be blocked, by placing a beam stop between the two
interferometric arms. The great advantage of the pulsed regime is that the
time of acceleration is the same for all atoms and, therefore, the grating
period $\lambda _{g}$ becomes velocity-independent. It allows one to relax
the severe requirements for longitudinal cooling. We will consider the
pulsed regime in more detail in a future publication.

In this article we have analyzed the role of the longitudinal degrees of
freedom using a perturbation theory in the scattering angle $\theta $. To
our knowledge only two other articles address this problem \cite{L1,L2} for
a finite value of $\theta $ in an atom interferometer consisting of a set of
spatially separated, resonant traveling waves. The consideration in \cite{L2}
that assumes the edges of the field envelopes are shorter than an atomic
de-Broglie wave length, while a quasiclassical approach was used in \cite{L1}%
. Notably, we failed to solve the relevant Schr\"{o}dinger equation for
scattering from a large angle beam splitter. However, we can stress that,
when seeking sinusoidal atom gratings of a period smaller than the optical
wavelength, but still larger than the de-Broglie wavelength, our
perturbation theory is sufficient. If $\theta \sim 1$ $\left( \lambda
_{dB}\sim \lambda \right) $ the atom wave packet will be broadened in
momentum space, which would destroy the sinusoidal shape of the atom grating
and defeat the purpose of our method.

\acknowledgments

We thank P. R. Berman and J. L. Cohen for help and recommendations, and T.
Chupp for discussion. This work is supported by the U. S. Office of Army
Research under Grant No. DAAD19-00-1-0412 and the National Science
Foundation under Grant No. PHY-9800981, Grant No. PHY-0098016, and by the
Office of the Vice President for Research and the College of Literature
Science and the Arts of the University of Michigan.
\end{multicols}

\appendix

\section{Ground state driving by $lin\bot lin$ polarized fields.}

\label{atom}

Consider an atom's interaction with a field consisting of two resonant waves
propagating along the $y$-axis, 
\begin{equation}
{\bf E}\left( {\bf r,}t\right) =\zeta \left( x,z\right) \sum_{j=1,2}%
{\textstyle{1 \over 2}}%
E_{j}{\bf e}_{j}\exp \left( -i\Omega _{j}t+iky\right) +c.c.,  \label{aa1}
\end{equation}
where $E_{j},{\bf e}_{j},\Omega _{j},k,\zeta \left( x,z\right) $ are the
field amplitude, polarization vector, frequency, wave vector, and envelope
function. The envelope function $\zeta \left( x,z\right) $ is the same for
both fields. For $lin\bot lin$ polarized fields one can choose ${\bf e}_{1}=%
{\bf \hat{z},}$ ${\bf e}_{2}={\bf \hat{x}}$.\ We assume that $\zeta \left(
x,z\right) $ is centered along the atomic beam trajectory, has a small width
along the $x$-axis $\left( \delta x\ll d\right) ,$ and a large width along $%
z $-axis $\left( \delta z\gg b\right) .$ The length scales $b$ and $d$ are
the atomic beam radius and the acceleration zone length, as, respectively,
defined in the our paper.

When field detunings from resonance are larger than the excited state decay
rate, the atomic ground state amplitudes $\psi _{Gm}$ ($G$ and $m$ are the
total moment and magnetic quantum number) evolve in the atomic rest frame $%
\left( x=ut\right) $ as\cite{Rmn} 
\begin{equation}
i\dot{\psi}_{G^{\prime }m^{\prime }}=\left\langle G^{\prime }m^{\prime
}\left| V\right| Gm\right\rangle \psi _{Gm},  \label{aa2}
\end{equation}
where 
\begin{mathletters}
\label{aa20}
\begin{eqnarray}
\left\langle G^{\prime }m^{\prime }\left| V\right| Gm\right\rangle
&=&\sum_{jj^{\prime }}\exp \left[ -i\delta _{G^{\prime }m^{\prime
},Gm}^{\left( jj^{\prime }\right) }t\right] A_{G^{\prime }G}^{\left(
jj^{\prime }\right) }\left( K\right) \left( -1\right) ^{G^{\prime }+m}\left( 
\begin{array}{ccc}
G^{\prime } & K & G \\ 
m^{\prime } & \nu _{k} & -m
\end{array}
\right) \varepsilon _{\nu _{k}}^{K}\left( jj^{\prime }\right) ,
\label{aa20a} \\
\delta _{G^{\prime }m_{g}^{\prime },Gm_{g}}^{\left( jj^{\prime }\right) }
&=&\Omega _{j}-\Omega _{j^{\prime }}-\omega _{G^{\prime }m^{\prime };Gm},
\label{aa20b} \\
A_{G^{\prime }G}^{\left( jj^{\prime }\right) }\left( K\right) &=&\left(
-1\right) ^{K+G^{\prime }+J_{G^{\prime }}+J_{G}+J_{H}+I}\left[ 
{\displaystyle{\chi _{J_{H}J_{G}}^{\left( j\right) }\left( \chi _{J_{H}J_{G}}^{\left( j^{\prime }\right) }\right) ^{\ast } \over \Delta _{J_{H},G}^{\left( j\right) }}}%
\right] \sqrt{\left( 2K+1\right) \left( 2G^{\prime }+1\right) \left(
2G+1\right) }  \nonumber \\
&&\times \left\{ 
\begin{array}{ccc}
J_{G} & J_{G^{\prime }} & K \\ 
1 & 1 & J_{H}
\end{array}
\right\} \left\{ 
\begin{array}{ccc}
J_{G^{\prime }} & I & G^{\prime } \\ 
G & K & J_{G}
\end{array}
\right\} ,  \label{aa20c} \\
\varepsilon _{\nu _{k}}^{K}\left( jj^{\prime }\right) &=&\left( -1\right)
^{\nu }e_{\nu }^{j}e_{-\nu ^{\prime }}^{j^{\prime }\ast }\sqrt{2K+1}\left( 
\begin{array}{ccc}
1 & 1 & K \\ 
\nu & \nu ^{\prime } & -\nu _{k}
\end{array}
\right) ,  \label{aa20d}
\end{eqnarray}
$\omega _{G^{\prime }m^{\prime },Gm}$ is the $G^{\prime }m^{\prime
}\rightarrow Gm$ transition frequency, $J_{G}$ and $J_{H}$ are the
electronic angular momenta of the ground and excited state manifolds, $I$ is
the nuclear spin, $\chi _{J_{H}J_{G}}^{\left( j\right) }=\left\langle
J_{H}\left| \left| d\right| \right| J_{G}\right\rangle E_{j}\zeta \left(
ut,0\right) /2\hbar $ is the Rabi frequency associated with field $j,$ $%
e_{\nu }^{j}$ is a spherical component of the polarization vector ${\bf e}%
_{j},$ $\left( ...\right) $ and $\left\{ ...\right\} $ are 3J- and
6J-symbols, and we have assumed that the detuning $\Delta _{J_{H},G}^{\left(
j\right) }$ is larger than the Zeeman and hyperfine splitting of the excited
states. To be specific, we put $J_{G}=1/2$ and $J_{H}=I=3/2,$ which
corresponds to the $D_{2}$ line in $^{\text{87}}$Rb.

Consider first the case where the fields are tuned to the two-photon
transitions between Zeeman sublevels of the $G=1$ manifold, i.e. $\Omega
_{2}=\Omega _{1}+\omega _{Z},$ where $\omega _{Z}=\omega _{11;10}=\omega
_{10;1,-1}.$ The wave function evolves as 
\end{mathletters}
\begin{mathletters}
\label{aa3}
\begin{eqnarray}
i\dot{\psi}_{11} &=&\chi _{0}\psi _{11}+\left( \chi -\chi ^{\ast
}e^{2i\omega _{Z}t}\right) \psi _{10,}  \label{aa3a} \\
i\dot{\psi}_{10} &=&\left( \chi ^{\ast }-\chi e^{-2i\omega _{Z}t}\right)
\psi _{11}+\left( \chi -\chi ^{\ast }e^{2i\omega _{Z}t}\right) \psi
_{1,-1}+\chi _{0}\psi _{10},  \label{aa3b} \\
i\dot{\psi}_{1,-1} &=&\left( \chi ^{\ast }-\chi e^{-2i\omega _{Z}t}\right)
\psi _{10}+\chi _{0}\psi _{1,-1},  \label{aa3c}
\end{eqnarray}
where $\chi _{0}=-3^{-1}\left[ A_{11}^{\left( 11\right) }\left( 0\right)
+A_{11}^{\left( 22\right) }\left( 0\right) \right] $ and $\chi
=2^{-3/2}3^{-1/2}A_{11}^{\left( 21\right) }\left( 1\right) $ are the
ac-Stark shift and effective Rabi frequency associated with transitions
between Zeeman sublevels. For a large Zeeman splitting, 
\end{mathletters}
\begin{equation}
\omega _{Z}\tau _{i}\gg 1,  \label{aa30}
\end{equation}
one neglects the rapidly oscillating terms in Eqs. (\ref{aa3}) to find the
wave functions after the field pulses: 
\begin{mathletters}
\label{aa4}
\begin{eqnarray}
\psi _{11}^{+} &=&\frac{1}{2}\exp \left[ -i\Lambda _{0}\right] \left\{ \left[
1+\cos \left( \Lambda \right) \right] \psi _{11}^{-}-\left[ 1-\cos \left(
\Lambda \right) \right] \psi _{1,-1}^{-}-i\sqrt{2}\sin \left( \Lambda
\right) \psi _{10}^{-}\right\} ,  \label{aa4a} \\
\psi _{10}^{+} &=&\exp \left[ -i\Lambda _{0}\right] \left\{ -i2^{-1/2}\sin
\left( \Lambda \right) \left( \psi _{11}^{-}+\psi _{-1}^{-}\right) +\cos
\Lambda \psi _{10}^{-}\right\} ,  \label{aa4b} \\
\psi _{1,-1}^{+} &=&\frac{1}{2}\exp \left[ -i\Lambda _{0}\right] \left[
-\left( 1-\cos \left( \Lambda \right) \right) \psi _{11}^{-}+\left( 1+\cos
\left( \Lambda \right) \right) \psi _{1,-1}^{-}-i\sqrt{2}\sin \left( \Lambda
\right) \psi _{10}^{-}\right] ,  \label{aa4c}
\end{eqnarray}
\end{mathletters}
where $\psi _{Gm}^{-}$ on the right hand side are the initial values of the
atomic wave function amplitudes, $\Lambda _{0}=\int_{-\infty }^{\infty
}dt_{1}\chi _{0}$, $\Lambda =2^{1/2}\int_{-\infty }^{\infty }dt_{1}\chi $
are field areas, and $\chi $ is assumed to be real. If $\left\{ \psi
_{11}^{-},\psi _{10}^{-},\psi _{1,-1}^{-}\right\} =\left\{ 0,1,0\right\} $,
one splits $100\%$ of the atoms between Zeeman sublevels $m=\pm 1$ using a $%
\frac{\pi }{2}$-pulse, 
\begin{equation}
\Lambda =\frac{\pi }{2}.  \label{aa5}
\end{equation}
For fields of Gaussian profile, 
\begin{equation}
\zeta \left( x,z\right) =\exp \left( -2\left( x/\delta x\right) ^{2}-2\left(
z/\delta z\right) ^{2}\right) ,  \label{aa6}
\end{equation}
one finds that the geometric average $P_{0}=\left( P_{1}P_{2}\right) ^{1/2}$
of the field powers $P_{1}$ and $P_{2}$ is given by 
\begin{equation}
P_{0}=\left[ \left( 2\pi \right) ^{3/2}\hbar m_{e}c^{2}/e^{2}\lambda
_{J_{H}J_{G}}f\left( J_{G},J_{H}\right) \right] u\delta z\Delta
_{J_{H},1}^{\left( 2\right) },  \label{aa7}
\end{equation}
where $m_{e}$ is the electron mass, $\lambda _{J_{H}J_{G}}$ and $f\left(
J_{G},J_{H}\right) $ are the wavelength and oscillator strength associated
with the excited-ground state transition.

Atoms in $\left| m=\pm 1\right\rangle $ sublevels start to accelerate in an
inhomogeneous magnetic field. To stop this acceleration at a later time, one
needs to return the atoms back to the $\left| m=0\right\rangle $ state.
Inserting initial conditions $\psi _{11}^{-}=1$ (or $\psi _{1,-1}^{-}=1$) in
Eqs. (\ref{aa4}) one sees that one can return at most half of the atoms to
the $\left| m=0\right\rangle $ state, again using a $\frac{\pi }{2}$-pulse.
The other half remains split between the $\left| m=\pm 1\right\rangle $
states.

This loss of atoms can not be avoided in the quadrupole $I$ (see Fig. \ref
{f03}), if one operates in the weak acceleration regime. However, for the
strong acceleration regime or for quadrupoles $II$ and $III,$ one can use
different fields along different arms of the interferometer and employ
another hyperfine sublevel to achieve a $100\%$ exchange between accelerated
and non-accelerated Zeeman sublevels.

For example, to start the acceleration in quadrupole $II,$ one chooses the
field frequency difference 
\begin{equation}
\Omega _{2}-\Omega _{1}=\omega _{21,10},  \label{aa8}
\end{equation}
such that under condition (\ref{aa30}), only the transition between
sublevels $\left| G=1,m=0\right\rangle $ and $\left| G=2,m=1\right\rangle $
occurs. The wave function amplitudes of this two-level system evolve as 
\begin{mathletters}
\label{aa9}
\begin{eqnarray}
i\dot{\psi}_{21} &=&\chi _{1}\psi _{21}+\chi \psi _{10},  \label{aa9a} \\
i\dot{\psi}_{10} &=&\chi \psi _{21}+\chi _{0}\psi _{10},  \label{aa9b}
\end{eqnarray}
where $\chi _{1}=15^{-1/2}\left( A_{22}^{\left( 11\right) }\left( 0\right)
+A_{22}^{\left( 22\right) }\left( 0\right) \right) ,$ $\chi
_{0}=-3^{-1}\left( A_{11}^{\left( 11\right) }\left( 0\right) +A_{11}^{\left(
22\right) }\left( 0\right) \right) ,$ $\chi =2^{-3/2}5^{-1/2}A_{21}^{\left(
21\right) }\left( 1\right) .$ To transfer all atoms between the sublevels,
one needs the ac-Stark shifts to be equal, $\chi _{1}=\chi _{0},$ which
means that the ratio of the fields' powers has to be chosen as 
\end{mathletters}
\begin{equation}
P_{2}/P_{1}=-1-2\omega _{21,10}/\Delta _{J_{H},1}^{\left( 1\right) }.
\label{aa10}
\end{equation}
This ratio is positive only for negative detuning, 
\[
-2\omega _{21,10}<\Delta _{J_{H},1}^{\left( 1\right) }<0. 
\]
To obtain a $100\%$ transfer between the levels, one should apply a $\pi $%
-pulse, for which $\int_{-\infty }^{\infty }dt\left| \chi \right| =\pi /2.$
This condition is an equation for the geometric average of the field powers.
Combining this equation with the Eq. (\ref{aa10}), one finds the powers, 
\begin{mathletters}
\label{aa11}
\begin{eqnarray}
P_{1} &=&P_{0}^{\prime }\left| 1+2\omega _{21,10}/\Delta _{J_{H},1}^{\left(
1\right) }\right| ^{-1/2},  \label{aa11a} \\
P_{2} &=&P_{0}^{\prime }\left| 1+2\omega _{21,10}/\Delta _{J_{H},1}^{\left(
1\right) }\right| ^{1/2},  \label{aa11b}
\end{eqnarray}
where 
\end{mathletters}
\begin{equation}
P_{0}^{\prime }=\left[ 4\pi ^{3/2}\hbar m_{e}c^{2}/3^{1/2}e^{2}\lambda
_{J_{H}J_{G}}f\left( J_{G},J_{H}\right) \right] u\delta z\Delta
_{J_{H},1}^{\left( 2\right) }.  \label{aa12}
\end{equation}

\section{Justification of the slowly varying amplitude approximation.}

\label{Justification}

In this Appendix we determine conditions under which it is valid to assume
that the amplitude $\psi \left( z,t\right) $ in Eq. (\ref{i51}) varies
slowly, i.e. the operator $Q$ given by Eqs. (\ref{i17}) leads to small
corrections to the zero-order approximation solution (\ref{i20}). To find
these conditions, one has to include the time dependence of the force and
the homogeneous part of the potential. At small times, $\left| t\right| \ll
\tau _{a},$ these are given by 
\begin{mathletters}
\label{a1}
\begin{eqnarray}
f\left( t\right) &\approx &f\left( 1+\xi t^{2}/\tau _{a}^{2}\right) ,
\label{a1a} \\
U_{0}\left( t\right) &=&U_{0}\left( 1+\nu t^{2}/\tau _{a}^{2}\right) ,
\label{a1b}
\end{eqnarray}
where $\nu \sim 1$ and $\xi $ is given by 
\end{mathletters}
\begin{equation}
\xi =2\left[ 3b_{s}^{2}\left( \alpha ^{2}-1\right) \left( \alpha
^{2}+1\right) ^{2}-16\alpha ^{2}\right] b_{s}^{-2}\left( \alpha
^{2}+1\right) ^{-4}.  \label{a2}
\end{equation}
for a potential produced by a magnetic quadrupole. We found that if (i) one
can neglect the first term in Eq. (\ref{i17a}) and (ii) approximation (\ref
{24}) is valid, then the slowly varying amplitude approximation is valid if
the correction (\ref{25}) is of a small relative weight $\left( \sim
\varepsilon _{q}\ll 1\right) .$ We now prove assumptions (i) and (ii).

We start from the first term in Eq. (\ref{i17a}). During the acceleration it
produces corrections to the wave function of relative weight $\int_{-\tau
}^{t}dt\frac{du}{dx}=\int_{-d}^{x}\frac{du\left( x\right) }{u\left( x\right) 
}\approx \frac{\Delta u}{u}$ and $\Delta u\hbar q/Mu^{3},$ where $\Delta u$
is a typical change of the velocity (\ref{i501}). Under condition (\ref{i50}%
), using the estimates $\frac{\Delta u}{u}\sim \left( U_{0}/Mu^{2}\right)
\delta ^{2},$ $U_{0}\sim fa,$ 
\begin{equation}
q\approx f\partial _{p_{0}}\sim fb/\hbar  \label{a21}
\end{equation}
and Eqs. (\ref{18}, \ref{19}), one finds that the weights are small $\left(
\sim \theta \delta \ll 1\text{ and }\theta ^{2}\beta \ll 1\right) ,$ and one
can eliminate the first term in Eq. (\ref{i17a}).

Now, consider the remaining part of Eq. (\ref{i17a}). For the operator $q$
one finds 
\begin{equation}
q=\partial _{t}-f\partial _{p_{0}}-i\left( \dot{\phi}-f\left( t\right)
\delta z/\hbar +p_{0}\delta \dot{z}/\hbar \right) .  \label{a3}
\end{equation}
Calculating $\dot{\phi}$ and $\delta z$ using Eqs. (\ref{i15}, \ref{a1}) one
finds 
\begin{eqnarray}
Q &=&-\frac{\hbar ^{2}}{2Mu^{2}}\left\{ \left( \partial _{t}-f\partial
_{p_{0}}\right) ^{2}+i\left( M\hbar \right) ^{-1}\left[ f\delta p-p_{0}f-\xi
f^{2}\tau ^{3}\tau _{a}^{-2}g^{\prime }\right. \right.  \nonumber \\
&&-\left. \left. 2\left( p_{0}\delta p+\xi f^{2}\tau ^{4}\tau
_{a}^{-2}g\right) \left( \partial _{t}-f\partial _{p_{0}}\right) \right]
-\left( M\hbar \right) ^{-2}\left( p_{0}\delta p+2\xi f^{2}\tau ^{4}\tau
_{a}^{-2}g\right) ^{2}\right\} ,  \label{a4}
\end{eqnarray}
where $g=\left[ -3\left( t/\tau \right) ^{4}-8\left( t/\tau \right)
^{3}-6\left( t/\tau \right) ^{2}+1\right] /12,$ $g^{\prime }=\tau \partial
g/\partial t.$

Since we have used the operator $Q$ above only for the evaluation of the
corrections, it is sufficient to consider the operator $Q$ acting only on
the unperturbed wave function $\Psi _{0}\left( p_{0},t\right) $. For weak
acceleration, when $\Psi _{0}\left( p_{0},t\right) =\Psi _{0}\left(
p_{0},-\tau \right) ,$ one can eliminate the time-derivative in Eq. (\ref{a4}%
). For strong acceleration, owing to the phase associated with the second
term in brackets in Eq. (\ref{36}), $\Psi _{0}\left( p_{0},t\right) =\exp %
\left[ -\frac{i}{\hbar }\int_{-\tau }^{t}dt_{1}U_{1}\left[ \delta z\left(
t_{1}\right) ,t_{1}\right] \right] \Psi _{0}\left( p_{0},-\tau \right) ,$
the time-derivative\ leads to the factor 
\begin{equation}
\partial _{t}\sim U_{1}\left[ \delta z\left( t\right) ,t\right] /\hbar .
\label{a5}
\end{equation}
The leading independent terms in the $U_{1}$ expansion (\ref{15b}) are
associated with the $\left( 2,m_{1}\right) $ and $\left( n_{2},0\right) $
elements in Table (\ref{17}). Retaining only these terms and assuming that $%
\delta z\left( t\right) \sim \Delta z$ and $t\sim \tau ,$ one obtains the
estimate, 
\begin{equation}
\partial _{t}\sim f\Delta z^{2}\left( \hbar a\right) ^{-1}\delta
^{2m_{1}}+f\Delta z^{n_{2}}\hbar ^{-1}a^{1-n_{2}}.  \label{a6}
\end{equation}
Comparing this result with estimate (\ref{a21}), one finds that the
time-derivative can still be eliminated even for strong acceleration if
parameters 
\begin{mathletters}
\label{a8}
\begin{eqnarray}
\eta _{3} &=&\theta ^{2}\beta ^{-1}\delta ^{2\left( m_{1}+1\right) },
\label{a8a} \\
\eta _{4} &=&\left( \delta \theta \right) ^{n_{2}}\beta ^{-1}  \label{a8b}
\end{eqnarray}
are small.

Assuming for estimates that $\delta p\sim f\tau ,$ one finds that the
weights of other contributions to $q^{2}$ differ from $\left( f\partial
_{p_{0}}\right) ^{2}$ by factors $\eta _{1},$ $\eta _{1}\eta _{2},$ $\delta
^{2}\eta _{1},$ $\eta _{5},$ $\eta _{1}^{2},$ $\eta _{1}^{2}\eta _{5,}$
where 
\end{mathletters}
\begin{equation}
\eta _{5}=\theta \delta ^{3}\beta ^{-1}.  \label{a9}
\end{equation}
From Table \ref{t1} in Sec. IV, one sees\ that all these factors are smaller
than unity, except the factor $\eta _{5}$, which is large for the strong
acceleration regime. The contribution of the order of $\eta _{5}$ is
proportional to the parameter $\xi .$ Since parameters $\xi $ and $c_{21}$
vanish simultaneously [compare Eqs. (\ref{162a}, \ref{a2})], one can exclude
in Eq. (\ref{a4}) all terms containing $\xi $ if the aspect ratio of the
quadrupole $\alpha $ and the relative bias field strength $b_{s}$ are chosen
to satisfy Eq. (\ref{32}). Therefore, for all cases under consideration, the
expression (\ref{24}) provides the main contribution to the correction
associated with a slowly varying amplitude approximation.

\end{document}